\documentclass{statsoc}

\usepackage[a4paper]{geometry}

\usepackage{amsmath,amssymb,amsthm,amsfonts}
\usepackage{rotating}
\usepackage{mathtools}

\usepackage{graphicx,psfrag,epsf}
\usepackage{enumerate}
\usepackage{natbib}
\usepackage{hyperref} 

\usepackage{ctable}
\usepackage{algpseudocode}

\newcommand{\R}{\mathbb{R}}
\newcommand{\E}{\mathbb{E}}

\DeclarePairedDelimiter\floor{\lfloor}{\rfloor}

\newcommand\T{\rule{0pt}{3.0ex}}       
\newcommand\Bot{\rule[-1.2ex]{0pt}{0pt}} 

\def\reals{\mathbb{R}}

\def\comp{\raise 1pt \hbox{$\scriptstyle\circ$}}

\def\argmin{\mathop{\rm argmin}\limits}
\def\argmax{\mathop{\rm argmax}\limits}

\def\upto{{\raise 1pt \hbox{$\scriptstyle \,\nearrow\,$}}}
\def\downto{{\raise 1pt \hbox{$\scriptstyle \,\searrow\,$}}}

\newtheorem{lemma}{Lemma}
\newtheorem{proposition}{Proposition}

\newtheorem{corollary}{Corollary}

\newtheorem{remark}{Remark}

\newtheorem*{assumption*}{\assumptionnumber}
\providecommand{\assumptionnumber}{}
\makeatletter
\newenvironment{assumption}[2]
{%
	\renewcommand{\assumptionnumber}{Assumption $\mathbf{#2}$#1}%
	\begin{assumption*}%
		\protected@edef\@currentlabel{$\mathbf{#2}$#1}%
	}
	{%
	\end{assumption*}
}

\usepackage[a4paper]{geometry}

\usepackage[utf8]{inputenc}

\usepackage{natbib}

\usepackage{color}
\usepackage{anyfontsize}
\usepackage{rotating}
\usepackage{ctable}
\usepackage{hyperref}

\usepackage{amsthm}
\usepackage{amssymb}

\usepackage{mathtools}
\usepackage{algpseudocode}

\usepackage{etoolbox}

\makeatletter
\patchcmd{\@makecaption}
{\parbox}
{\advance\@tempdima-\fontdimen2} 
{}{}
\makeatother  

\makeatletter
\def\@textbottom{\vskip \z@ \@plus 1pt}
\let\@texttop\relax
\makeatother

\def\reals{{\mathbb{R}}}

\def\E{{\mathbb E}}

\def\R{{\mathbb R}}

\def\reals{\mathbb{R}}

\def\comp{\raise 1pt \hbox{$\scriptstyle\circ$}}

\def\argmin{\mathop{\rm argmin}\limits}
\def\argmax{\mathop{\rm argmax}\limits}

\def\upto{{\raise 1pt \hbox{$\scriptstyle \,\nearrow\,$}}}
\def\downto{{\raise 1pt \hbox{$\scriptstyle \,\searrow\,$}}}

\makeatother

\title[Non-parametric Structural Change Detection]{Non-parametric Structural
        Change Detection in Multivariate Systems}

\author[Pekka Malo {\it et al.}]{Pekka Malo} 
\address{Department of Information and Service Management, Aalto University,
        P.O. Box 21220, FI-00076 Aalto, Finland} 
\email{pekka.malo@aalto.fi} 

\author[Pekka Malo {\it et al.}]{Lauri Viitasaari}
\address{Department of Mathematics and Statistics,
        University of Helsinki, P.O. Box 68, FI-00014 Helsinki, Finland}
\email{lauri.viitasaari@helsinki.fi}

\author[Pekka Malo {\it et al.}]{Olga Gorskikh}
\address{Department of Information and Service Management, Aalto University,
        P.O. Box 21220, FI-00076 Aalto, Finland}  
\email{olga.gorskikh@aalto.fi}

\author[Pekka Malo {\it et al.}]{Pauliina Ilmonen}
\address{Department of Mathematics and Systems Analysis, Aalto University,
        P.O. Box 11100, FI-00076 Aalto, Finland}
\email{pauliina.ilmonen@aalto.fi}

\begin{document}

\begin{abstract}
        Structural change detection problems are often encountered in analytics
        and econometrics, where the performance of a model can be significantly affected
        by unforeseen changes in the underlying relationships. Although these problems
        have a comparatively long history in statistics, the number of studies done
        in the context of multivariate data under nonparametric settings is still
        small. In this paper, we propose a consistent method for detecting multiple
        structural changes in a system of related regressions over a large dimensional
        variable space. In most applications, practitioners also do not have a priori
        information on the relevance of different variables, and therefore, both
        locations of structural changes as well as the corresponding sparse regression
        coefficients need to be estimated simultaneously. The method combines nonparametric
        energy distance minimization principle with penalized regression techniques.
        After showing asymptotic consistency of the model, we compare the proposed
        approach with competing methods in a simulation study. As an example of a
        large scale application, we consider structural change point detection in
        the context of news analytics during the recent financial crisis period.
\end{abstract}

\section{Introduction}\label{sec:intro}

Interest towards large dimensional multivariate regression and interdependence
analysis has surged due to their relevance for mining predictive relationships
out of massive data sets~\citep{yuan07,negahban-wainwright-2011}. Many of
these problems are characterized by the dual challenge of learning several
related models simultaneously while allowing them to account for a large
pool of candidate variables that are partly shared across the individual
relationships~\citep{abernethy09,negahban-wainwright-2011,agarwal12}. Such
large dimensional modeling tasks are commonly encountered in practical applications,
such as financial forecasting, news analytics, or marketing, where the objective
is to predict the development of many possibly related indicators simultaneously~\citep{stock-watson09,groen13,fan11}.
However, a further layer of complexity is introduced, when the underlying
predictive relationships are recognized to undergo multiple structural changes
when longer time periods are considered~\citep{qian_su_2016,chopin06,bai-perron98,bai-perron03}.
For example, in marketing applications, it is rational to expect that consumer
preferences can change rapidly in response to major product or technological
innovations. Often, the practitioner also does not have a priori information
on the relevance of the candidate variables, and therefore it becomes natural
to let the data decide which variables should be retained. When
combined with the requirement of detecting an unknown number of change points
in multivariate data, encountering the simultaneous variable selection problem
limits the applicability of earlier methods, which assume either a fixed
and typically very small set of contributing explanatory variables~\citep{li-perron-17,qu-perron-07}
or investigate only single equation models~\citep{bai-perron03}.   

In this paper, we consider large dimensional regression problems, where the
objective is to estimate a collection of related regressions 
over a varying set of features while allowing the model to be exposed to
multiple
structural changes. A structural change is defined as a point that separates
a time-ordered sample into two parts having different linear structures.
Throughout, we treat both number as well as locations of the structural change
points as unknown variables. We also assume that the model structure is sparse
and that the potential structural changes take place in a discontinuous manner,
where both parameter estimates as well as the number of variables with non-zero
coefficients can vary from one regime to another. Further, we do not make
any assumptions regarding the underlying distribution beyond the requirement
of very weak moment conditions on the regressors and residuals. Since this
kind of problem is typically ill-posed due to the dimensionality concerns,
it is natural to impose sparsity constraints or regularization on the problem.
Regularization is formulated as a convex optimization problem consisting
of a loss term and a regularizer. The framework of this paper works under
very general requirements for the admissible regularizers as well as loss
functions. 

Our paper has two main objectives. The first is to propose a non-parametric
method that can consistently estimate an unknown number of structural change
points in a large dimensional multivariate linear regression model. To avoid
imposing distributional assumptions, we approach the problem using an energy
distance framework that is based on $U$-statistics~\citep{rizzo-szekely16,
        szekely-rizzo13,szekely-rizzo14,szekely-rizzo05}. The asymptotic results
are obtained under quite general conditions. The second objective is to look
at the problem from algorithm-design perspective, and ensure that the estimation
principle can be implemented in a computationally efficient manner. To address
this, two algorithms are suggested. The first is based on the principle of
dynamic programming, which has been successfully applied also in the earlier
literature by~\cite{bai-perron03}. This approach gives a consistent way to
obtain the global minimizers of energy distance statistic. However, it remains
computationally quite demanding, and requires $O(T^2)$ operations for any
given number of structural change points. The second algorithm is a more
efficient heuristic with performance of order $O(T)$ but with no guarantee
of finding the global minimizers. However, our extensive simulation study
gives evidence on its ability to detect the structural changes with an accuracy
that is on par with the dynamic programming principle. Therefore, it can
be a preferred choice for practitioners dealing with large models and long
time periods that usually have many structural changes. As an example, we
consider structural change detection in the context of news analytics.

Though change point analysis has attracted widespread attention across different fields \citep{cho-fryzlewicz-15}, the literature on structural change detection,
especially in the context of systems of multivariate equations, has remained
relatively sparse~\citep{li-perron-17,qu-perron-07,kurozumi-azai-07}. Whereas
change point
analysis commonly refers to detection of breaks in trend or distributional
changes (e.g., shift in mean or variance) in univariate or multivariate series
~\citep{Ruggieri2016, matteson-james14,harchaoui-10}, structural change analysis
is focused on detecting changes in the underlying predictive relationship~\citep{bai-perron03,
        qu-perron-07,qian_su_2016,li-perron-17}. Hence, along with changes in distribution
or
trend, breaks can be attributed to shifts in the model parameters or changes
in the pool of relevant explanatory variables. Although, the two lines of
research, change point analysis and structural change analysis, have evolved
simultaneously, their development has been driven by different fields of
study. While change point analysis (or data segmentation~\citep{fryzlewicz-14}),
has been directly motivated by applications in signal processing and bioinformatics,
structural change analysis is popular in social disciplines, business and
economics~\citep{bai-perron03,qu-perron-07,qian_su_2016}.  

Another important distinction in literature is made between parametric and
nonparametric setups. In parametric change point analysis, the underlying
distributions are assumed to belong to some known family that admits use
of log-likelihood functions in the analysis~\citep{davis-06,lebarbier-05,lavielle-06}.
Recently, non-parametric methods have gained traction as they are considered
applicable to  a wider range of applications~\citep{matteson-james14,hariz-07}.
However, many of these approaches require estimation of density functions
or density ratios~\citep{kawahara-12,kanamori-09,liu-13}. Also rank statistics
and energy distance statistics have been considered~\citep{matteson-james14}.
One of the key benefits of energy statistics is their simplicity. Since they
are based on Euclidean distances~\citep{szekely-rizzo05}, the energy statistics
are easy to compute also in multivariate settings. However, it is noted that
these nonparametric approaches have been proposed in the context of change
point analysis to detect distributional changes rather than structural breaks. In this paper, we show how the idea of using energy distance statistics for distributional change detection~\citep{matteson-james14} can be extended to structural change detection in models with large number of potential explanatory variables.

Against this backdrop, we propose a new nonparametric method for detecting
structural changes in multivariate data. In comparison to the literature,
our work differs in three aspects. First, we allow the modeling to take place
with large pool of candidate variables, and acknowledge that each structural
change can be accompanied by change in the collection of variables with non-zero
coefficients, which is quite different from the settings in~\cite{bai-perron03,qu-perron-07}
and \cite{li-perron-17}. Also, unlike~\cite{qian_su_2016} who employ group
fused lasso penalty to detect change point locations, we use sparsity constraints
to guide variable selection within regimes rather than to detect the regime
boundaries. Based on the experiments, our approach appears to produce more
parsimonious models in terms of the number of change points. Second, the
use of nonparametric energy statistics allows us to relax important distributional
assumptions. In particular, this has the benefit of reducing sensitivity
towards outliers and fat-tailed residual distributions. Finally, differing
from most of the prior work, our method
is designed to handle change point detection in multivariate systems of equations
rather than restricting to a single predictive relationship~\citep{qian_su_2016,bai-perron03}.

The rest of the paper is organized as follows. In Section 2, we present the
model and the estimation principle based on minimization of energy distances.
Section 3 discusses definitions and properties of energy distance statistics.
Section 4 presents assumptions and the asymptotic consistency results for
the model. This is followed by description of nonparametric goodness-of-fit
statistics in Section 5, which are then used to guide the algorithms are
outlined in Section 6. In Section 7, we show the results from computational
studies, where our approach applied to simulated and real data. As an example
of    a large scale problem, we consider structural change detection in the
context of business news analytics, where the objective is to understand
how different types of financial news events are reflected  company valuations.
Concluding remarks are given in Section 8.

\section{Model}
Consider the following multiple regression model with $k$ change points ($k+1$
regimes):
\[
y_t = x_t'\beta_j + u_t, \quad t=T_{j-1}+1, \dots, T_j
\]
for $j=1,\dots,k+1$. By convention we have that $T_0=0$ and $T_{k+1}=T$.
In this model, $y_t \in \reals^q $ denotes an observed independent variable,
$u_t \in \reals^q$ is the
disturbance, $x_t \in \reals^p$ is a vector of covariates, and $\beta_j \in
\R^{p\times q}$
$(j=1,\dots,k+1)$ are the corresponding matrix of coefficients. Throughout
the paper we denote by $|\cdot|$ the Euclidean norm and by $\Vert \cdot \Vert$
the corresponding operator norm for the matrices. Note that the norms depend
on the dimensions which we have omitted on the notation. The sequence
of unknown break points are denoted by indices $(T_1,\dots,T_k)$. The purpose
is to estimate the unknown regression coefficients and the change points
based on the observed data $(y_t, x_t)$. Throughout the paper, we denote
the true value of a parameter with a 0 superscript. In particular, the true
values for coefficients and the change points are denoted by $\beta^0=(\beta_1^0,\dots,\beta_{k+1}^0)$
and $(T_1^0,\dots,T_k^0)$, respectively. In general, the number of change
points can be assumed to be an unknown variable with true value of $k^0$.
However, to simplify our discussion on the general estimation principles,
we will for now treat the number of change points $k$ as known. Methods for
estimating $k$ will be presented in later parts of the paper. 

\begin{figure}[h!]
        \centering
        \includegraphics[width=14cm,height=13.9cm]{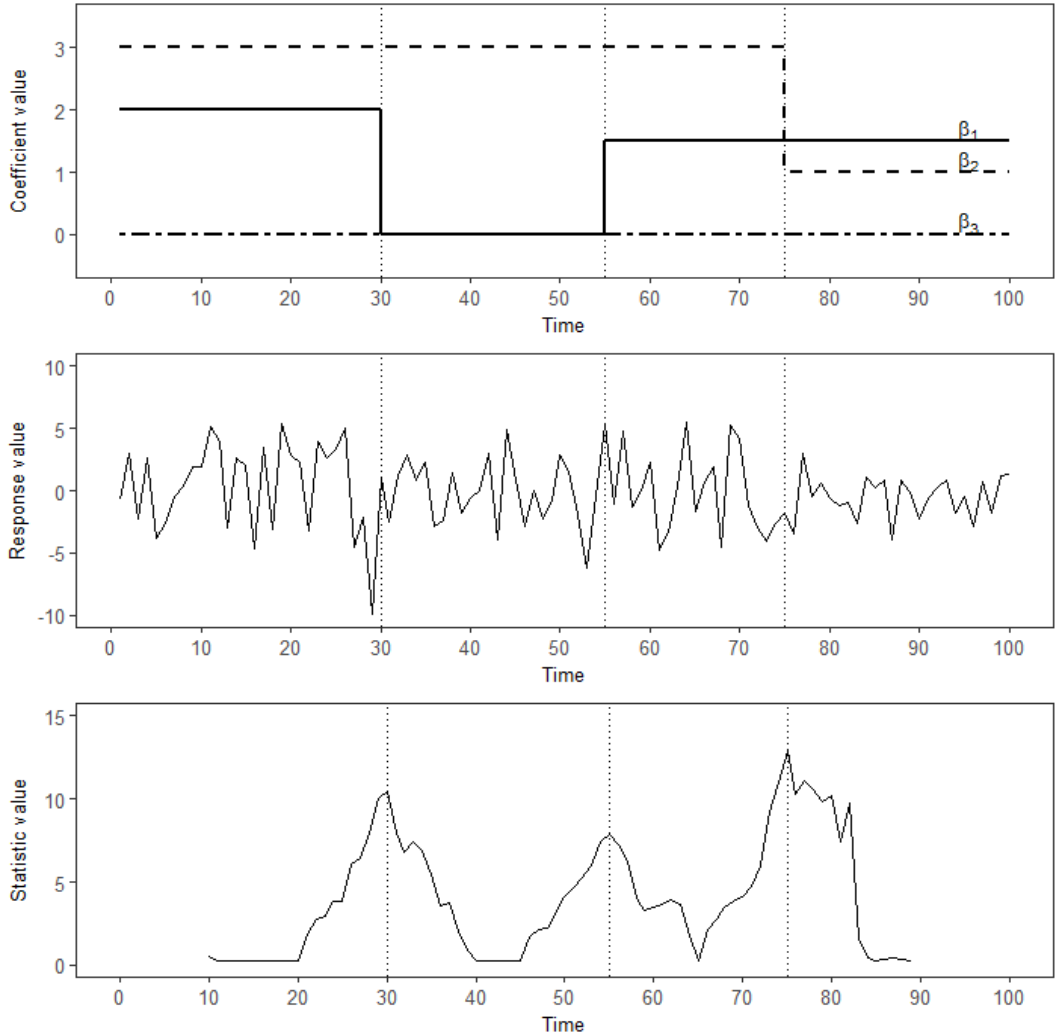}
        \caption{\small Energy-distance based detection of structural changes
                in a single equation model with 4 regimes and 3 variables. Locations of the
                structural changes are highlighted by dashed vertical lines. The first graph
                shows the evolution of coefficients over time and second graph shows the
                time series for the response variable. The third graph shows the corresponding
                changes in the energy-distance measure, which is used to detect the regime
                boundaries.}\label{fig:betas}
\end{figure}   

The estimation method is constructed as a hybrid of penalized regression
technique and non-parametric testing strategy. We assume that the coefficients
$\{\beta_j\}$ representing different regimes exhibit sparsity such that the
effective number of non-zero coefficients in each $\beta_j$ is less than
$p$. The large number of potential covariates motivates the use of regularization
techniques. Given a $k$-partitioning $\{T_j\}=(T_1,\dots,T_k)$, the estimates
of $\beta$ are obtained as minimizers of the empirical risk  
\begin{equation}\label{eq:lasso}
(\hat{\beta}_1,\dots,\hat{\beta}_{k+1}) = \argmin_{\beta} \mathcal{L}_T(\beta;
\{T_j\}) = \sum_{j=1}^{k+1} \sum_{t= T_{i-1}+1}^{T_i} \psi(x_t'\beta_j -
y_t) + \gamma_n \varphi(\beta),
\end{equation}
where $\psi$ is a strictly convex loss function and $\varphi$ is a convex
function such that both $\psi$ and $\varphi$ attain their global minimums
at zero. To highlight the dependence on the partitioning, the penalized estimates
are denoted by $\hat{\beta}(\{T_j\})$. Substituting these into the model
equation gives us estimates of the regression residuals. Let $\mathcal{U}=\{U_1,\dots,U_{k+1}\}$
represent the partitioning of the regression residuals $\hat{u}_t$ into clusters
such that $U_j=\{\hat{u}_{T_{j-1}+1},\dots, \hat{u}_{T_j}\}$.  The change
points are then defined as global minimizers of the goodness-of-fit statistic
\begin{equation}\label{eq:min-energy}
(\hat{T}_1,\dots,\hat{T}_k) = \argmin_{T_1,\dots,T_k} \sum_{1 \leq i < j\leq
        k+1} \left(\frac{n_i+n_j}{2T}\right) d_{\alpha}(U_i,U_j),
\end{equation}
where $n_i$ and $n_j$ denote the sample sizes of $U_i$ and $U_j$, respectively.
The minimization is taken over all partitions of the timeline such that $T_j-T_{j-1}\geq
\varepsilon T$ for some $\varepsilon >0$. The function $d_{\alpha}$ is a
measure of the empirical distance between the distributions of the partitioned
disturbances by \cite{szekely-rizzo13,szekely-rizzo14,szekely-rizzo05}. Here
the objective is to detect the change points such that the partitioned
model residuals $\mathcal{U}$ can be interpreted as $k+1$ random samples
from distributions with cumulative distribution function $F_1,\dots,F_{k+1}$,
for which the null hypothesis of equal distributions $H_0: F_1 = \cdots
= F_{k+1}$ holds. The test is implemented as a bootstrap statistic, which
is discussed in Section~\ref{sec:nonparametric-tests}. 

A stylized example of the approach is given in Figure~\ref{fig:betas}, which
shows functioning of the model in a single equation example with only three
variables and four regimes, i.e. $q=1$, $p=3$ and $k=3$. The number of non-zero
variables can change in any regime and not all candidate variables need to
contribute to the relation. In this example,
the model residuals and explanatory variables are all normally distributed.
However, as shown by our experiments, the relative benefits of our model
are mainly realized in large dimensional settings, where the normality assumption
is not met due to presence of outliers or fat-tailed residuals.  These are
the circumstances, where the use of non-parametric energy-distance becomes
helpful. To further motivate our approach, we will in next section
discuss
the key properties of $d_{\alpha}$ and introduce the notion of energy distance
as a non-parametric measure of dispersion that can be computed based on Euclidean
distances between all pairs of sample elements.

\begin{remark}
        Occurrence of a structural change does not necessarily imply a distributional change in the joint distribution of $(y_t, x_t)$. A simple example could  be constructed using a model with only two explanatory variables and a single response, where the explanatory variables follow the same distribution. Then a time point, where the coefficients of the two explanatory variables get exchanged, would not count as a distributional change point, but would still be considered a structural change point that should be detected.
\end{remark}

\section{Energy distance}

Energy distance is a metric that measures the distance between the distributions
of random vectors, which was introduced and popularized by~\cite{rizzo-szekely16,szekely-rizzo13,szekely-rizzo14,szekely-rizzo05}.
The energy distance is zero if an only if the distributions are identical,
otherwise it will diverge. The notion derives from the concept of Newton's
potential energy by considering statistical observations as objects in a
metric space that are governed by statistical potential energy. Since its
introduction the energy distance and the more general class of energy statistics
have been utilized in a number of applications ranging from testing independence
by distance covariance to non-parametric tests for equality of distributions.
Our study as well as the e-divisive algorithm by~\cite{matteson-james14}
show how energy distance can be utilized for analysis of change points or
structural breaks in time series data.

\subsection{Energy distance for two samples}\label{sec:energy-distance}

As proven by Sz\'ekely and Rizzo, it can be shown that energy distance satisfies
all axioms of a metric, and therefore it provides a characterization of equality
of distributions as well as a theoretical basis for development of multivariate
analysis based on Euclidean distances.

\begin{lemma}\label{lemma:energy}
        Suppose $u,u'\stackrel{iid}{\sim} F_u$ and $v,v'\stackrel{iid}{\sim} F_v$,
        and that $u,u',v$, and $v'$ are mutually independent random variables in
        $\reals^d$. If $E(|u|^{\alpha}+|v|^{\alpha})<\infty$, for any $\alpha \in(0,2)$,
        then the characteristic function based divergence measure between distributions
        can be defined based on Euclidean distances as
        $$
        \mathcal{E}(u,v;\alpha)=2E|u-v|^{\alpha}-E|u-u'|^{\alpha}-E|v-v'|^{\alpha}
        $$
        such that $\mathcal{E}(u,v;\alpha)=0$ if and only if $u$ and $v$ are identically
        distributed.
\end{lemma}
The corresponding empirical divergence measure can then be defined in the
spirit of $U$-statistics. If $V_n=\{v_i : i=1,\dots,n\}$ and $W_m=\{w_j :
j=1,\dots,m\}$ are independent iid samples from distributions $F_v$ and $F_w$,
such that $E|v|^{\alpha},E|w|^{\alpha} < \infty$, we can use the divergence
to define the empirical energy distance measure as
\begin{equation}
\begin{split}
d_{\alpha}(V_n,W_m) & =\frac{mn}{m+n}\hat{\mathcal{E}}(V_n,W_m;\alpha)\\
&=\frac{mn}{m+n} (2\hat{\mu}^{\alpha}_{vw}-\hat{\mu}^{\alpha}_v-\hat{\mu}^{\alpha}_w),
\end{split}
\end{equation}
where 
\[
\hat{\mu}^{\alpha}_{vw} =\frac{1}{mn}\sum_{i=1}^n\sum_{j=1}^m |v_i-w_j|^{\alpha},
\quad \hat{\mu}^{\alpha}_v  = \binom{n}{2}^{-1}\sum_{1 \leq i < k \leq n}|v_i-v_k|^{\alpha},
\quad \text{and}  \]
\[
\hat{\mu}^{\alpha}_w  = \binom{m}{2}^{-1}\sum_{1 \leq j < k \leq m}|w_j-w_k|^{\alpha}.
\]
This empirical measure is based on Euclidean distances between sample elements
and is $O(\max{(m^2,n^2)})$. Under the given assumptions, the strong law
of large numbers for U-statistics~\cite{hoeffding61} and continuity theorem
imply that $\hat{\mathcal{E}}(V_n,W_m;\alpha) \to \mathcal{E}(v,w;\alpha)$
almost surely as $n,m\to\infty$. When equal distributions are assumed, the
energy distance measure $d_{\alpha}(V_n,W_n)$ convergences to a non-degenerate
random variable. Conversely, if the distributions are unequal, it follows
that the energy distance diverges, i.e. $d_{\alpha}(V_n,W_m)\to\infty$ almost
surely as $n,m\to\infty$, since $\mathcal{E}(v,w;\alpha)>0$ for unequal distributions.

\subsection{Multi-sample energy distance} 

For any partitioning $\{T_j\}$, let $S_{\alpha}$ denote the objective function
in~\eqref{eq:min-energy}, i.e.
\begin{equation}\label{eq:multisample-energy}
S_{\alpha}(U_1,\dots,U_{k+1})=\sum_{1 \leq i < j\leq
        k+1} \left(\frac{n_i+n_j}{2T}\right) d_{\alpha}(U_i,U_j)
\end{equation}
where $U_j=\{\hat{u}_{T_{j-1}+1},\dots, \hat{u}_{T_j}\}$ is the sequence
of residuals from regime $j$. As seen from the following corollary of Lemma~\ref{lemma:energy},
$S_{\alpha}$ statistic can be viewed as a multi-sample extension of
the two-sample distance measure introduced in Section~\ref{sec:energy-distance}.
\begin{corollary}\label{lemma:multisample-energy}
        For all $p$-dimensional samples $U_1,\dots,U_{k+1}$, $k\geq 1$, and $\alpha
        \in
        (0,2)$, the following statements hold: (i) $S_{\alpha}(U_1,\dots,U_{k+1})\geq
        0$; and (ii) $S_{\alpha}(U_1,\dots,U_{k+1}) = 0$ if and only if $U_1,\dots,
        U_{k+1}$
        are equally distributed.
\end{corollary}
The proof of the result is obtained by applying induction argument on Lemma
1. It is clear from the construction that the statistic is likely to share
many interesting similarities with ANOVA. By interpreting $S_{\alpha}$ as
a multi-sample test of equal distributions, it can be considered as a type
of generalization of the hypothesis of equal means. In fact, as shown by
\cite{rizzo-szekely10}, the connection to analysis of variance can be obtained
through the special case $\alpha=2$, when the $d_2$-distance for a univariate
response variable measures variance.

\section{Consistency}

In this section, we study the consistency of the estimated change point fractions
in the case of a single change point as well as the generalization of the
result into the case of multiple change points. We denote the estimated change
point fractions and their corresponding true values by $\hat{\lambda}=(\hat{\lambda}_1,\dots,\hat{\lambda}_k)=(\hat{T}_1/T,\dots,\hat{T}_k/T)$
and $\lambda^0=(\lambda_1^0,\dots,\lambda_k^0)$, respectively.

Throughout the following discussions on the statistical properties of the
estimators, we will rely on the following assumptions. 

\begin{assumption}{1}{A}\label{as:1}
        The change points are asymptotically distinct such that $T_i^0=[T\lambda_i^0]$,
        where $0 < \lambda_1^0 < \cdots < \lambda_k^0 < 1$ and $\lambda_0^0 = 0$
        and $\lambda_{k+1}^0=1$. 
\end{assumption}

\begin{assumption}{2}{A}\label{as:2}
        The model regressors $x_t$ are identically distributed within regions, i.e.
        $x_t \sim X_k$ for every $T_{k-1}+1\leq t \leq T_k$. Furthermore we have,
        for a given $\alpha \in(0,2)$, that $E(|x_t|^{2\alpha})<\infty$.
\end{assumption}

\begin{assumption}{3}{A}\label{as:3}
        The model disturbances $u_t$ are independent and identically distributed.
        Further, the disturbances $u_t$ are assumed to be independent of the regressors
        $x_s$ for all $t$ and $s$.
        Finally, we assume that, for a given $\alpha \in(0,2)$, we have $E(|u_t|^{2\alpha})<\infty$.
\end{assumption}

\begin{assumption}{4}{A}\label{as:lasso}
        For any given change points $T_i,i=1,2,\ldots,k$, the regularized estimators
        $\hat\beta_i$ converges in probability to some constant $\beta^c_i$. That
        is, we have
        $
        \Vert \hat\beta_i - \beta^c_i\Vert \rightarrow 0
        $
        in probability.
        Moreover, for any $i$,
        the regularized estimator $\hat\beta_i$ is consistent only if $T_i^0\leq
        T_i < T_{i+1}\leq T_{i+1}^0$.
\end{assumption}

\begin{assumption}{5}{A}\label{as:technical}
        Let $c_1,c_2\in \R^{p\times q}$ be arbitrary matrices and let
        $$
        Y_{i,j} = |u_i-u_j + x_i'c_1 - x_j'c_2|^\alpha,
        $$
        where $T_{k-1}+1 \leq i,j \leq T_k$ for some $k$.
        We assume that the regressors $x_s$ are asymptotically independent in the
        sense that, as
        $$
        \min_{l,j\in\{1,2,3,4\},l\neq j}|i_l-i_j| \rightarrow \infty,
        $$
        we have
        $$
        Cov(Y_{i_1,i_2},Y_{i_3,i_4}) \rightarrow 0
        $$
        and
        $$
        \E[Y_{i_1,i_2}] \rightarrow \E|U-\tilde{U} + X_k'c_1 - \tilde{X}_k'c_2|^\alpha,
        $$
        where $U,\tilde{U}$ are independent copies of $u$ and $X',\tilde{X}'$ are
        independent copies of $x$.
\end{assumption}

The first technical assumption is very natural. Indeed, if the change points
are not (asymptotically) distinct, then one may simply remove one. The second
and the third technical assumptions give moment conditions and distributional
assumptions for the regressors and for the disturbance terms that guarantee
the convergence of the empirical energy distances.

The fourth technical assumption is also a natural one. The first statement
of the assumption means that, as the number of observations increase, the
regularized estimators converge to some constants. We emphasize the fact
that the constants might be, and usually are, wrong ones, unless the change
points are estimated correctly. Moreover, the consistency assumption states
that the regularized estimators are consistent if the estimation is based
on the observations lying on the correct intervals.

The fifth technical assumption is used to guarantee the convergence of the
empirical energy distances to some constant quantities. This assumption simply
states that the regressors are asymptotically independent. (Assumption that
is widely used in the literature). The intuition behind this assumption is
that, as the number of observations increase on every subinterval, one can
think that the dependence of the regressors between fixed time points  is
spread among the time points in the middle. 

Note that these assumptions are quite mild. Typical vector autoregressive
models, for example, fulfill the above assumptions A1-A5.

\subsection{Single change point}

In order to
obtain consistency, we apply the following elementary lemma providing a version
of weak law of large numbers for weakly dependent double arrays.

\begin{lemma}
        \label{lma:weak_law}
        Let $X_{i_1,i_2}, i_1,i_2=1,\ldots,n$ denote a double array of random variables  with 
$$
\sup_{i_1,i_2} \E(X^2_{i_1,i_2}) < \infty.
$$ 
Assume there exists a constant
        $C\in \R$ such that 
        $$
        \E\left[(X_{i_1,i_2}-C)(X_{i_3,i_4}-C)\right] \rightarrow 0
        $$
        as $\min_{k,j\in\{1,2,3,4\},k\neq j}|i_k-i_j| \rightarrow \infty.$
        Then, as $n\rightarrow \infty$, $\frac{1}{n^2}\sum_{i_1,i_2=1}^n X_{i_1,i_2}
        \rightarrow C$ in probability.
\end{lemma}

The proof of Lemma~\ref{lma:weak_law} is provided in Appendix A.1. Now together with assumptions A1-A5, we can show the consistency of the estimator in the case of a single change point.

\begin{proposition}\label{proposition1}
        Let $\hat{T}_1$ denote the estimated energy-distance minimizing change point
        location, as defined in equation \eqref{eq:min-energy}. Suppose that $\frac{\hat{T}_1}{T}$,
        as $T\to\infty$, converges in probability to $\hat{\lambda}_1 \in(0,1)$.
        Then, under A1-A5, we have $\hat{\lambda}_1 =\lambda_1^0$.
\end{proposition}

The proof  is obtained by contradiction (see Appendix A.2 for detailed proof). Assume that $\lambda_1^0$ is not consistently
estimated, i.e. $\hat{\lambda}_1 \neq \lambda_1^0$. Without loss of generality,
we assume that the estimated change
point $\hat{T}_1$ satisfies $\hat{T}_1 < T\lambda_1^0$, giving
us a partitioning $I_1=[0, \hat{T}_1]$, $I_2=[\hat{T}_1+1, T_1^0]$ and $I_3=[T_1^0+1,T]$.

We denote by $|A|$ the size of a set $A$. In particular, we have that $|I_1
\cup
I_2 \cup I_3| = T$, $|I_3|=(1-\lambda_1^0)T$, $|I_1| \sim \hat{\lambda}_1T$,
$|I_2|\sim(\lambda_1^0 -\hat{\lambda}_1)T$, and $|I_2 \cup I_3| \sim (1-\hat{\lambda}_1)T$
(where the notation $f(T) \sim g(T)$ refers to the usual interpretation $\lim_{T\to\infty}
\frac{f(T)}{g(T)} = 1$). For
notational simplicity, we denote by $\hat{\beta}(i)$ the estimator corresponding
to the region where $i$ belongs. That is, for $\hat{\beta}_i,i=1,2$ denoting
the regularized estimates, we have $\hat\beta(i) = \hat{\beta}_1$ for all
$i \in I_1$, and 
$\hat{\beta}(i) = \hat{\beta}_2$ for all $i \in I_2 \cup I_3$. Similarly,
we denote by $\beta^0(i)$ the correct value corresponding to the region where
$i$ belongs. That is, as the true change point is $T_1^0$, we have $\beta^0(i)
= \beta^0_1$ for all $i \in I_1 \cup I_2$ and $\beta^0(i) = \beta^0_2$ for
all $i\in I_3$. We also denote by $\beta(i)$ and $\beta^c_k$ the limits related
to Assumption 4. More precisely, we always have $\hat\beta(i)
\rightarrow \beta(i)$ and $\hat\beta_k \rightarrow \beta^c_k$. Moreover,
we have  
$\hat\beta(i) = \hat\beta_1\rightarrow \beta^c_1=\beta^0_1$ for all $i \in
I_1$, as region $I_1$ is a subset of the correct interval $[0,T_1^0]$. For
$i \in I_2 \cup I_3$, we have $\hat{\beta}(i) = \hat{\beta}_2 \rightarrow
\beta^c_2$, and thus $\beta(i) = \beta^c_2$.  

Denote by $\hat{u}_i = u_i - x_i'(\hat{\beta}(i)-\beta^0(i)),i\in I_k,k=1,2,3$
the corresponding estimated residuals and let $U_1 = \{\hat{u}_t\}_{t\in
        I_1}$ and $U_2 = \{\hat{u}_t\}_{t\in
        I_2\cup I_3}$ denote the collections of regularized residuals from different
intervals. We set
\[
\hat{\mu}^{\alpha}_{U_1,U_2}  =\frac{1}{|I_1||I_2\cup I_3|}\sum_{j \in I_1,
        i \in I_2\cup I_3} |\hat{u}_i-\hat{u}_j|^{\alpha}, \hat{\mu}^{\alpha}_{U_1}
= \frac{1}{|I_1|^2}\sum_{i,j \in I_1}|\hat{u}_i-\hat{u}_j|^{\alpha},
\]
and
\[
\hat{\mu}^{\alpha}_{U_2}  = \frac{1}{|I_2\cup I_3|^2}\sum_{i,j\in I_2\cup
        I_3}|\hat{u}_i-\hat{u}_j|^{\alpha}.
\]
We prove that 
\begin{equation}
\label{eq:empirical_limit}
2\hat{\mu}^{\alpha}_{U_1,U_2}-\hat{\mu}^{\alpha}_{U_1}-\hat{\mu}^{\alpha}_{U_2}
\rightarrow C>0,
\end{equation}
where $C$ is a constant and the convergence holds in probability. From this
we get
$$
d_{\alpha}(U_1, U_2) = \frac{|I_1||I_2\cup I_3|}{2|I_1|+2|I_2\cup I_3|}\left(2\hat{\mu}^{\alpha}_{U_1,U_2}-\hat{\mu}^{\alpha}_{U_1}-\hat{\mu}^{\alpha}_{U_2}\right)
\rightarrow \infty.
$$
Consequently, $\hat{T}_1$ cannot be a minimizer for  the model equation 2
of Section 2 (in the paper),
which leads to the expected contradiction. 

We divide the rest of the proof into three steps. In step 1 we consider the
differences $\hat{u}_i - \hat{u}_j$ that depend on the entire data set. In
step 2 we calculate the limits of the terms $\hat{\mu}^{\alpha}_{U_1,U_2}$,
$\hat{\mu}^{\alpha}_{U_1}$, and $\hat{\mu}^{\alpha}_{U_2}$. Finally, in step
3, we show \eqref{eq:empirical_limit}. For complete technical proof, see Appendix A.2.

\subsection{Multiple change points}

Having established consistency in the case of a single change point, it is now straightforward to extend the result to the case of multiple change points.

\begin{proposition}
        Let $\hat{T}_j,j=1,\ldots,k$ denote the estimated energy-distance minimizing
        change point locations, as defined in equation \eqref{eq:min-energy}. Suppose
        that, for all $j=1,\ldots,k$, as $T\to\infty$, the quantity $\frac{\hat{T}_j}{T}$
        converges in probability to $\hat{\lambda}_j \in(0,1)$. 
        Then, under A1-A5, $\hat{\lambda}_j = \lambda_j^0$ for all $j=1,\ldots,k$.
\end{proposition}

The proof is again obtained by assuming contradiction and using Proposition 1 together with Lemma 2. The technical details of the proof are provided in Appendix A.3.

\section{Non-parametric change point tests}\label{sec:nonparametric-tests}

The estimates from \eqref{eq:lasso} and \eqref{eq:min-energy} are consistent
when the number of actual change points is known. However, in practice, the
number of true change point is generally not known. Therefore, in order to
construct a suitable algorithm, we need a test statistic that allows us to
check whether the proposed partitioning produces an acceptable fit.

\subsection{Goodness of fit test for $k$ change point model}\label{sec:gof}

Let $(T_1,\dots,T_k)$ be any hypothesized sequence of change points, and
let $(U_1,\dots,U_{k+1})$, $U_i\sim F_i$, denote the corresponding sequences
of model residuals for the $k+1$ regimes. To test for homogeneity in distribution,
\begin{equation}\label{hypothesis}
H_0: F_1 = \cdots = F_{k+1}, k\geq 1, 
\end{equation}
versus the composite alternative $F_i \neq F_j$ for some $1\leq i < j \leq
k+1$, we can apply the distance components statistic by~\cite{rizzo-szekely10}.
If $H_0$ is rejected, we conclude that there is at least one change point
that has not been identified.

The test statistic is constructed in a manner analogous to ANOVA, and it
is based on the following decomposition theorem that is obtained by direct
application of the results in~\cite{rizzo-szekely10} on the $k$ change point
problem. Define the total dispersion of the estimated regime residuals as
\begin{equation}
T_{\alpha}(U_1,\dots,U_{k+1})=\frac{T}{2}\mu_{\alpha}(U,U),
\end{equation}
where $U=\sum_{j=1}^{k+1}U_j$ is the pooled sample of regime residuals, and
\[
\mu_{\alpha}(A,B)=\frac{1}{n_1 n_2}\sum_{i=1}^{n_1}\sum_{j=1}^{n_2}|a_i-b_j|^{\alpha}
\]
for any sets $A$ and $B$ of size $n_1$ and $n_2$, respectively. Similarly,
we can define the within-sample dispersion statistic as
\begin{equation}
W_{\alpha}(U_1,\dots,U_{k+1})=\sum_{j=1}^{k+1}\frac{n_j}{2}\mu_{\alpha}(U_j,
U_j)
\end{equation}
\begin{proposition}
        For $k\geq 1$, the total dispersion $T_{\alpha}$ of the $(k+1)$-regime residuals
        can be decomposed as
        \begin{equation}
        T_{\alpha}(U_1,\dots,U_{k+1})=S_{\alpha}(U_1,\dots,U_{k+1})+W_{\alpha}(U_1,\dots,U_{k+1}),
        \end{equation}  
        where $S_{\alpha}$ is the multi-sample energy distance \eqref{eq:multisample-energy}
        and $W_{\alpha}$ is the within-sample dispersion. If $\alpha \in (0,2)$,
        then the test statistic,
        \begin{equation}\label{eq:goodness-of-fit}
        F_{\alpha}(U_1,\dots,U_{k+1}) = \frac{S_{\alpha}(U_1,\dots,U_{k+1})/k}{W_{\alpha}(U_1,\dots,U_{k+1})/(T-k-1)},
        \end{equation}
        for hypothesis~\eqref{hypothesis} is statistically consistent against all
        alternatives with finite second moments. 
\end{proposition}

The test for \eqref{hypothesis} can be implemented as a permutation test.
To ensure computational tractability of the procedure, we approximate the
p-value by performing a sequence of $R$ random permutations. The permutation
test can be used as a stopping criterion for the estimation procedures discussed
in the subsequent sections. Let $\mathcal{T}$ be a vector of indices in the
pooled sample of residuals, $U=\{u_t\}_{t\in \mathcal{T}}$. With slight abuse
of notation, we define statistic $F_{\alpha}(U;\pi)$ as $F_{\alpha}(u_{\pi(\mathcal{T})})$,
where $\pi(\mathcal{T})$ is a permutation of the elements in $\mathcal{T}$.
If the null hypothesis holds, then the statistics $F_{\alpha}(u_t)$ and $F_{\alpha}(u_{\pi(\mathcal{T})})$
are identically distributed for every permutation of $\mathcal{T}$. The permutation
test procedure is implemented as follows. First, compute the test statistic
$F_{\alpha}=F_{\alpha}(U; \mathcal{T})$. Next, for each permutation $\pi_r
= \pi(\mathcal{T})$, $r\in\{1,\dots,R\}$, compute the statistic $F_{\alpha,r}=F_{\alpha}(U;\pi_r)$.
The approximate $p$-value is then defined as $\#\{r : F_{\alpha,r} \geq F_{\alpha}\}/(R+1)$.

\subsection{Specific change-point location test}\label{sec:loc-test}

The above goodness-of-fit statistic can also be used to construct a test
for evaluating a given change-point location. Suppose our current model has
$k$ correctly identified change points. Let $\delta \in\Delta_{j,\eta}$ be
a proposed specific location for a new change point within $j$th regime,
where
\begin{equation}\label{eq:delta-set}
\Delta_{j,\eta} = \left\{t : \hat{T}_{j-1}+(\hat{T}_j-\hat{T}_{j-1})\eta
\leq t \leq \hat{T}_j - (\hat{T}_j-\hat{T}_{j-1})\eta \right\}
\end{equation} 
is a subinterval within $j$th regime with $\eta>0$ large enough to ensure
sufficiency of data around the hypothesized change point location. This allows
us to define segments $D^{\delta-}_1 = \{(y_t,x_t) : t\in (\hat{T}_{j-1},\delta)\}$
and $D^{\delta+}_2 = \{(y_t,x_t) : t\in [\delta,\hat{T}_j)\}$, which divide
the current $j$th regime into left and right parts.

Under the null hypothesis of no change at $\delta$, we can estimate a model
on $D^{\delta-}_1$ to obtain post-regularization coefficients $\tilde{\beta}^{\delta-}$
and the corresponding residuals $U^{\delta-}_{1}=\{\tilde{u}_t = y_t - x_t'\tilde{\beta}^{\delta-}
: (y_t,x_t)\in D^{\delta-}_1\}$. Since no change is assumed to take place,
the coefficients $\tilde{\beta}^{\delta-}$ estimated from the left segment
can also be applied on  $D^{\delta+}_2$ to produce residuals $U^{\delta+}_{2|1}=\{\tilde{u}_t
= y_t - x_t'\tilde{\beta}^{\delta-}
: (y_t,x_t)\in D^{\delta+}_2\}$ for the right segment. The fact that we reuse
the coefficients estimated from the first segment is highlighted by the subscript.

Now a test statistic for the null of no change point at $\delta$ is obtained
by considering a test for homogeneity in distribution, 
\begin{equation}\label{hypothesis}
H_0: F^{\delta-} = F^{\delta+},  
\end{equation}
where $U^{\delta-}_1\sim F^{\delta-}$ and $U^{\delta+}_{2|1}\sim F^{\delta+}$.
For any $\alpha\in (0,2)$, the corresponding test statistic is then given
by $F_{\alpha}(U^{\delta-}_1,U^{\delta+}_{2|1})$ as defined in Section~\ref{sec:gof}.
Again, in the absence of distributional assumptions, this statistic can be
implemented as a permutation test. 
\section{Computing the global minimizers}

A brute-force approach to solve the minimization problem defined by \eqref{eq:lasso}
and \eqref{eq:min-energy} is to consider a grid search. As the number of
change point is a discrete parameter that can take only a finite number of
values, use of grid search would guarantee the detection of optimal break
points. However, as the number of potential change point increases $k>2$,
the strategy will quickly become inefficient as the number of operations
required would increase at rate $O(T^k)$. As proposed by~\cite{bai-perron03},
this can be alleviated by considering a strategy that is motivated by the
principle of dynamic programming~\citep{bellman-roth69,fisher58}. The approach
suggested in this section is somewhat similar, but the special nature of
the non-parametric test statistic and the use of regularization estimation
makes the problem computationally more demanding and increase the need for
memory.  

\subsection{Estimation with a known number of change points}\label{sec:known-change}

Let $S_{\alpha}^{\star}(\{T_{m,n}\})$ denote the value of the multi-sample
energy-distance~\eqref{eq:multisample-energy} obtained from the optimal partitioning
of the first $n$ observations using $m$ change points. The optimal partitioning
can be expressed as a solution for a recursive problem:
\begin{equation}\label{recursive-problem}
S_{\alpha}^{\star}(\{T_{k,T}\}) = \inf_{k\tau \leq t \leq T-\tau}\left[S_{\alpha}^{\star}(\{T_{k-1,t}\})
+ A_{\alpha}(t+1,T) \right],
\end{equation}
where $A_{\alpha}$ is the additional energy distance produced by adding the
residuals estimated from period $t+1$ to $T$, and $\tau >0$ is an imposed
constraint on the minimum length of any regime. If $U_1^{\star},\dots,U_k^{\star}$
represent the $k$ residual samples
that follow from the optimal partitioning of $t$ first observations with
$k-1$ change
points, and $U_{k+1}$ denotes from $t+1$ to $T$, the additional energy distance
is given by
\begin{equation}
\begin{split}
A_{\alpha}(t+1,T) &= S_{\alpha}(U_1^{\star},\dots,U_{k}^{\star}, U_{k+1})
- S_{\alpha}(U_1^{\star},\dots,U_{k}^{\star}) \\
&= \sum_{j=1}^{k} \left(\frac{n_j+n_{k+1}}{2T}\right) d_{\alpha}(U_j^{\star},U_{k+1})
\end{split}
\end{equation}
where $n_j$ and $n_{k+1}$ are the sample sizes of $U_j^{\star}$ and $U_{k+1}$.

The solution approach is based on the fact that the number of possible regimes
is at most $T(T+1)/2$ and hence the number of times the regularized estimation
needs to be performed is no more than of order $O(T^2)$. Furthermore, it
is important to note that many of these candidate regimes are not admissible,
when we take into account the requirement that the minimum admissible length
for any regime considered by the model is $\tau$. Majority of the cost of
this algorithm follows from the computation of a triangular matrix of pairwise
energy distances between all admissible regimes. Once the distance matrix
is known, the recursive formulation~\eqref{recursive-problem} to find the
optimal $k$-partitioning can be solved quickly, and will essentially follow
the approach suggested in~\cite{bai-perron03}. For a given $k\geq 1$, the
recursive algorithm is outlined as follows: 

{\it Step 1}: Start by finding the optimal single change point partitions
for all sub-samples that allow a potential change point to occur in $[\tau,
T-k\tau]$. This will require storage of $T-(k+1)\tau+1$ single point partitions
and the residuals corresponding to these models. The ending dates for the
partitions will be in $[2\tau, T-(k-1)\tau]$. 

{\it Step 2}: The second step will proceed by computing optimal partitions
with two change points that have ending dates in $[3\tau,T-(k-2)\tau]$. For
each possible ending date, we will then find which single change point partition
from the first step will minimize the total energy distance of thus obtained
two change point partitioning. To avoid duplicate computation of energy distances,
any pairwise distance computation between potential segments should be stored
for later use. As a result, we get a collection of $T-(k+1)\tau+1$ models
with two change points. 

{\it Step 3}: The steps will continue in a sequential manner until a set
of $T-(k+1)\tau +1$ models with optimal $k-1$ partitions is obtained, where
ending dates are in range $[(k-1)\tau, T-2\tau]$. The algorithm will now
terminate by finding which of these $k-1$ partitions will minimize the energy
distance of the complete sequence, and hence produce a solution for \eqref{recursive-problem}.

\subsection{Estimation with an unknown number of change points}\label{sec:unknown-change}

Generally, the number of change points is not known apriori, and needs to
be estimated along with the locations of the breaks. To address this, we
suggest complementing the above dynamic programming approach with a sequence
of nonparametric change point tests that can be used as a termination criterion.

Let $p_0$ be a selected critical value for tests, and let $M_k$ be an initial
model with $k$ change points (small number) that has been estimated with
the approach described in Section~\ref{sec:known-change}. In the spirit of
Section~\ref{sec:loc-test}, we can now consider an approach, where each of
the $k+1$ segments in the $k$ change point model is evaluated for an additional
change point. If the model with additional change point has considerably
smaller multi-sample energy statistic than the $k$ point model, we can conclude
in favor of the model $k+1$ points.  Suppose that $\hat{T}_1,\dots,\hat{T}_k$
is a sequence of estimated $k$ change points that globally minimize the multi-sample
energy distance in a sequence of $T$ observations. The ideal location for
the new structural change point is found by solving 
\begin{equation}\label{eq:sequential-test}
\min_{1\leq j \leq k+1} \ \min_{\delta \in \Delta_{j,\eta}} T_{\alpha}(U_1,\dots,U_{j-1},
U^{\delta-}_{j,1}, U^{\delta+}_{j,2}, \dots,U_{k+1}),
\end{equation}
where $\Delta_{j,\eta}$ is defined as in~\eqref{eq:delta-set}, and $U^{\delta-}_{j,1}$,
$U^{\delta+}_{j,2}$ denote the residuals associated with the new partitioning
at $\delta$, respectively.


%
%


\subsection{Nonparametric splitting algorithm}\label{sec:NSA} 

Solving the recursive problem~\eqref{recursive-problem} requires $O(T^2)$
operations for any $k$. To provide a faster alternative, we can consider
a heuristic that gives similar results under most settings. The logic of
the algorithm resembles the structure of binary segmentation, but instead
of marking the segment boundaries as direct estimates for change points,
we use the splitting technique to only zoom into promising regions without
making any statements on the exact locations of the change points at this
stage. The determination of the exact change point locations is done at the
last stage. 

Again, we can use pseudo-code to describe the procedure as follows.
Let $s$ and $e$ done the start and end points of the timeline, where the
change points are expected to occur. We require that $\tau  \leq s < e \leq
T-\tau$, where $\tau$ is the minimum length of regime. The parameter $l
\geq \tau$ controls the number of segments used in the initial search, and
$p_0$ is the selected critical value to be used for the test statistics.
The last parameter $\gamma$ controls the rate at which the size of search
regions at different stages is reduced. 

\begin{small}
        \begin{algorithmic}
                \Function{NSA}{$s$, $e$, $l$, $p_0$, $\gamma$}
                \State $R:=\floor*{\frac{e-s}{l}}$ is the number of segments with minimum
                length of $l$
                \State $\mathcal{I}^R_{s,e}:=$ ordered partitioning of interval $[s,e]$
                into $R$ segments $[s_r,e_r]$,  
                \State  $r=1,\dots,R$ (of equal length). 
                \If{first time of calling the method}
                \State Augment sets $[s_0,e_0]:=[0,s]$ and $[s_{R+1},e_{R+1}]:=[e,T]$ into
                $\mathcal{I}^R_{s,e}$.
                \EndIf
                \For{\textbf{each} $[s_r,e_r] \in \mathcal{I}^R_{s,e}$}
                \State Define $D_r:=\{(y_t,x_t) : t\in [s_r,e_r]\}$ and $D_{r+1}:=\{(y_t,x_t)
                : t\in [s_{r+1},e_{r+1}]\}$.
                \State Estimate regularized coefficients $\beta_r $ on $D_r$.
                \State Compute residuals $U_r$ and $U_{r+1}$ using $\beta_r$ on both data-segments.
                \State Calculate $p_r:=$ bootstrap $p$-value for $F_{\alpha}(U_r,U_{r+1})$
                \If{$p_r < p_0$ and $e_{r+1}-s_r > 2\tau$}
                \State Drill down into the region $D_r\cup D_{r+1}$:    
                \State \Call{NSA}{$s_r-\tau$, $e_{r+1}+\tau$, $l'$, $p_0$} with   $l':=\gamma
                l \geq \tau$ (e.g., $\gamma=0.5$). 
                \ElsIf{$p_r < p_0$}
                \State Find the exact change point location in $\tau$-extended region:
                \State $\delta^{\star}:=\argmax_{\delta\in\{s_r-\tau,\dots, e_{r+1}+\tau\}}
                F_{\alpha}(U^{\delta-}_1,U^{\delta+}_{2|1})$
                \State add $\delta^{\star}$ to the set of estimated change points
                \EndIf 
                \EndFor
                \EndFunction
        \end{algorithmic}
\end{small}

The procedure is initialized by calling NSA with $s$ and $e$ corresponding
to the maximum admissible interval. The choice of $l$ gives an upper bound
of $\floor*{T/l}$ for the expected number of change points. The main approach
used in NSA is to sequentially split the overall timeline into smaller segments.
The splitting is continued only in those regions that are indicated by distributional
homogeneity test statistics as areas where a structural change may have occurred.
As in Section~\ref{sec:loc-test},  the comparison of each pair of regions
is carried out under the null hypothesis of no change; i.e., the coefficients
estimated from the first region are assumed to be valid also on the second.
It is noteworthy that these steps are only limiting the potential search
regions without trying to actually locate the change points. 

Once the search has narrowed down into small enough regions, we can start
to locate the exact change-points. To account for the fact that the change
point may occur at the boundaries of the region, we expand the region by
adding a $\tau$-neighborhood for every candidate point. If $[s_r, e_{r+1}]$
represents the final search region, its $\tau$-expansion is given by $[s_r-\tau,e_{r+1}+\tau]$.
This will now allow for the former region boundaries $s_r$ and $e_{r+1}$
to be also considered as possible locations for structural change.

\section{Simulation studies}

In this section, we compare the performance of the energy distance based
approaches against the leading competitors that are available as R packages
or  as source code from authors. Since the main benefit of DP (Dynamic Programming, Sections~\ref{sec:known-change}-\ref{sec:unknown-change}) and NSA (Nonparametric Splitting Algorithm, Section~\ref{sec:NSA}) is their
ability to operate even under heavy-tailed errors or outlier contamination,
we construct several test data sets to get insights on the circumstances where different algorithms should be used. 

\subsection{Simulation settings}

In the experiments, we consider a data generating process with three structural breaks. Both univariate as well as multivariate response variables are considered, $y_t\in \reals^q$. The explanatory variables  $x_t \in \reals^p$ are assumed to be i.i.d. and follow a p-dimensional normal distribution, i.e. \(x_t \sim N_p(0,1)\). The error terms are assumed to be i.i.d. with distribution $u_t\sim F$, where $F$ is either a normal distribution $N(0,0.1)$ or Student's t-distribution $t(3)$. As a result, we have 
\[
y_t = \sum_{j=1}^{4}{x_t'\beta_jI(t\in[T_{j-1},T_j])} + u_t + \nu_t, \quad t=T_{j-1}+1, \dots, T_j,
\]
where \(T_0=0, \ T_1=60, \ T_2 = 300, \ T_3=480, \ T_4=600\). The extra term \(\nu_t\) represents an outlier that takes values from distribution $N(0,10)$ with probability $p_o$ and is zero otherwise, i.e. $\nu_t \sim p_0 N(0,10) + (1-p_0)\delta_0$. Using this framework, we have created 10 models, which differ in terms of the error distribution, amount of outliers, and number of explanatory variables. The model configurations for the experiments with univariate $(y_t\in \reals)$ and multivariate $(y_t\in\reals^3)$ responses are given in Tables~\ref{tab:univariate-coefs} and~\ref{tab:multivariate-coefs}, respectively. Models (1)-(8) are low-dimensional with 5 explanatory variables, while models (9) and (10) are high-dimensional with 100 explanatory variables.  However, only a subset of the variables is contributing within each regime. 
\begin{table}
        \caption{\label{tab:univariate-coefs}Summary of the univariate models used in the simulation study. Here $p$ is the number of explanatory variables, $d$ describes the magnitude of the change between subsequent regimes, $F$ is the error distribution and $p_0$ is the proportion of outliers.}        
        \centering                                                                                                                      
        \begin{small}                                                                                                                            
                                                                                        \fbox{                                                    
                \begin{tabular}{l p{0.5cm} p{0.5cm} p{1.6cm} p{1.6cm} p{1.6cm} p{1.6cm} p{1cm} p{1cm}} \hline\T

                        {\bf Model}     & $p$ & $d$ &   $\beta_1$       &       $\beta_2$       &       $\beta_3$       &       $\beta_4$       &       {$F$}   &       {$p_o$} \\\hline\T                      
                        (1)     &       5 & 1 & (1,1,1,0,0)     &       (2,1,1,0,0)     &       (1,1,1,0,0)     &       (1,2,1,0,0)     &       N(0,0.1)        &       0\%     \\                      
                        (2)     &       5 &1 & (1,1,1,0,0)      &       (2,1,1,0,0)     &       (1,1,1,0,0)     &       (1,2,1,0,0)     &       t(3)    &       0\%     \\                      
                        (3)     &       5 &1 & (1,1,1,0,0)      &       (2,1,1,0,0)     &       (1,1,1,0,0)     &       (1,2,1,0,0)     &       N(0,0.1)        &       10\%    \\                      
                        (4)     &       5 &1 & (1,1,1,0,0)      &       (2,1,1,0,0)     &       (1,1,1,0,0)     &       (1,2,1,0,0)     &       t(3)    &       10\%    \\\hline\T                      
                        (5)     &       5 &2 & (1,1,1,0,0)      &       (1,3,1,0,0)     &       (3,3,1,0,0)     &       (5,3,1,0,0)     &       N(0,0.1)        &       0\%     \\                      
                        (6)     &       5 &2 & (1,1,1,0,0)      &       (1,3,1,0,0)     &       (3,3,1,0,0)     &       (5,3,1,0,0)     &       t(3)    &       0\%     \\                      
                        (7)     &       5 &2 & (1,1,1,0,0)      &       (1,3,1,0,0)     &       (3,3,1,0,0)     &       (5,3,1,0,0)     &       N(0,0.1)        &       10\%    \\                      
                        (8)     &       5 &2 & (1,1,1,0,0)      &       (1,3,1,0,0)     &       (3,3,1,0,0)     &       (5,3,1,0,0)     &       t(3)    &       10\%    \\\hline\T                              
                        (9)     & 100 &2 
                        & $\beta_{17,j}$=1, $\beta_{42,j}$=1, $\beta_{67,j}$=1, $\beta_{88,j}$=1, $\beta_{91,j}$=1, $\beta_{i,j}$=0 for all other $i$ 
                        & $\beta_{17,j}$=1, $\beta_{42,j}$=3, $\beta_{67,j}$=1, $\beta_{88,j}$=1, $\beta_{91,j}$=1, $\beta_{i,j}$=0 for all other $i$
                        & $\beta_{17,j}$=3, $\beta_{42,j}$=3, $\beta_{67,j}$=1, $\beta_{88,j}$=1, $\beta_{91,j}$=1, $\beta_{i,j}$=0 for all other $i$   
                        & $\beta_{17,j}$=5, $\beta_{42,j}$=3, $\beta_{67,j}$=1, $\beta_{88,j}$=1, $\beta_{91,j}$=1, $\beta_{i,j}$=0 for all other $i$
                        &       N(0,0.1)        &       0\%     \\      \hline\T
                        (10)&   100 &2 
                        & $\beta_{17,j}$=1, $\beta_{42,j}$=1, $\beta_{67,j}$=1, $\beta_{88,j}$=1, $\beta_{91,j}$=1, $\beta_{i,j}$=0 for all other $i$ 
                        & $\beta_{17,j}$=1, $\beta_{42,j}$=3, $\beta_{67,j}$=1, $\beta_{88,j}$=1, $\beta_{91,j}$=1, $\beta_{i,j}$=0 for all other $i$
                        & $\beta_{17,j}$=3, $\beta_{42,j}$=3, $\beta_{67,j}$=1, $\beta_{88,j}$=1, $\beta_{91,j}$=1, $\beta_{i,j}$=0 for all other $i$   
                        & $\beta_{17,j}$=5, $\beta_{42,j}$=3, $\beta_{67,j}$=1, $\beta_{88,j}$=1, $\beta_{91,j}$=1, $\beta_{i,j}$=0 for all other $i$
                        &       N(0,0.1)        &       10\%    \Bot \\                 
                        \hline                                                                                                                          
                \end{tabular} 
                }                                                                                                                          
        \end{small}                                                                                                                              
\end{table}

\begin{table}     
 \caption{\label{tab:multivariate-coefs}Summary of the multivariate models
used in the simulation study. Here $p$ is the number of explanatory variables,
$d$ describes the magnitude of the change between subsequent regimes, $F$
is the error distribution and $p_0$ is the proportion of outliers.}                                                                                                                           
        \centering                                                                                                                      
        \begin{scriptsize}                                                                                                                              \fbox{
               
                \begin{tabular}{l p{0.5cm} p{0.5cm} p{0.5cm} p{1.5cm} p{1.5cm} p{1.5cm} p{1.5cm} p{1cm} p{1cm}} \hline\T

                        {\bf Model}     & $p$ & $d$ &  $y_t^j$ &        $\beta_1$       &       $\beta_2$       &       $\beta_3$       &       $\beta_4$       &       {$F$}   &       {$p_o$} \\\hline\T                      
                        (1)     &       5 & 1 & $y_t^1$ &(1,1,1,0,0)    &       (2,1,1,0,0)     &       (2,1,1,0,0)     &       (2,1,1,0,0)     &       N(0,0.1)        &       0\%     \\                      
                        &         &   & $y_t^2$ &(2,1,1,0,0)    &       (2,1,1,0,0)     &       (1,1,1,0,0)     &       (1,1,1,0,0)     &                        &               \\
                        &         &   & $y_t^3$ &(1,1,1,0,0)    &       (1,1,1,0,0)     &       (1,1,1,0,0)     &       (1,2,1,0,0)     &                &               \\\hline\T
                        (2)     &       5 & 1 & $y_t^1$ &(1,1,1,0,0)    &       (2,1,1,0,0)     &       (2,1,1,0,0)     &       (2,1,1,0,0)     &       t(3)    &       0\%     \\                      
                        &         &   & $y_t^2$ &(2,1,1,0,0)    &       (2,1,1,0,0)     &       (1,1,1,0,0)     &       (1,1,1,0,0)     &                        &               \\
                        &         &   & $y_t^3$ &(1,1,1,0,0)    &       (1,1,1,0,0)     &       (1,1,1,0,0)     &       (1,2,1,0,0)     &                &               \\\hline\T              
                        (3)     &       5 & 1 & $y_t^1$ &(1,1,1,0,0)    &       (2,1,1,0,0)     &       (2,1,1,0,0)     &       (2,1,1,0,0)     &       N(0,0.1)        &       10\%    \\                      
                        &         &   & $y_t^2$ &(2,1,1,0,0)    &       (2,1,1,0,0)     &       (1,1,1,0,0)     &       (1,1,1,0,0)     &                        &               \\
                        &         &   & $y_t^3$ &(1,1,1,0,0)    &       (1,1,1,0,0)     &       (1,1,1,0,0)     &       (1,2,1,0,0)     &                &               \\\hline\T
                        (4)     &       5 & 1 & $y_t^1$ &(1,1,1,0,0)    &       (2,1,1,0,0)     &       (2,1,1,0,0)     &       (2,1,1,0,0)     &       t(3)    &       10\%    \\                      
                        &         &   & $y_t^2$ &(2,1,1,0,0)    &       (2,1,1,0,0)     &       (1,1,1,0,0)     &       (1,1,1,0,0)     &                        &               \\
                        &         &   & $y_t^3$ &(1,1,1,0,0)    &       (1,1,1,0,0)     &       (1,1,1,0,0)     &       (1,2,1,0,0)     &                &               \\\hline\T
                        (5)     &       5 & 2 & $y_t^1$& (1,1,1,0,0)    &       (1,3,1,0,0)     &       (1,3,1,0,0)     &       (5,3,1,0,0)     &       N(0,0.1)        &       0\%     \\                      
                        &         &   & $y_t^2$ &(1,3,1,0,0)    &       (1,3,1,0,0)     &       (3,3,1,0,0)     &       (1,3,1,0,0)     &                        &               \\
                        &         &   & $y_t^3$ &(3,3,1,0,0)    &       (3,3,1,0,0)     &       (3,3,1,0,0)     &       (3,3,1,0,0)     &                &               \\\hline\T
                        (6)     &       5 & 2 & $y_t^1$& (1,1,1,0,0)    &       (1,3,1,0,0)     &       (1,3,1,0,0)     &       (5,3,1,0,0)     &       t(3)    &       0\%     \\                      
                        &         &   & $y_t^2$ &(1,3,1,0,0)    &       (1,3,1,0,0)     &       (3,3,1,0,0)     &       (1,3,1,0,0)     &                        &               \\
                        &         &   & $y_t^3$ &(3,3,1,0,0)    &       (3,3,1,0,0)     &       (3,3,1,0,0)     &       (3,3,1,0,0)     &                &               \\\hline\T
                        (7)     &       5 & 2 & $y_t^1$& (1,1,1,0,0)    &       (1,3,1,0,0)     &       (1,3,1,0,0)     &       (5,3,1,0,0)     &       N(0,0.1)        &       10\%    \\                      
                        &         &   & $y_t^2$ &(1,3,1,0,0)    &       (1,3,1,0,0)     &       (3,3,1,0,0)     &       (1,3,1,0,0)     &                        &               \\
                        &         &   & $y_t^3$ &(3,3,1,0,0)    &       (3,3,1,0,0)     &       (3,3,1,0,0)     &       (3,3,1,0,0)     &                &               \\\hline\T
                        (8)     &       5 & 2 & $y_t^1$& (1,1,1,0,0)    &       (1,3,1,0,0)     &       (1,3,1,0,0)     &       (5,3,1,0,0)     &       t(3)    &       10\%    \\                      
                        &         &   & $y_t^2$ &(1,3,1,0,0)    &       (1,3,1,0,0)     &       (3,3,1,0,0)     &       (1,3,1,0,0)     &                        &               \\
                        &         &   & $y_t^3$ &(3,3,1,0,0)    &       (3,3,1,0,0)     &       (3,3,1,0,0)     &       (3,3,1,0,0)     &                &               \\\hline\T
                        (9)     &100  & 2 & $y_t^1$
                        & $\beta_{17,j}$=1, $\beta_{42,j}$=1, $\beta_{67,j}$=1, $\beta_{88,j}$=1, $\beta_{91,j}$=1, $\beta_{i,j}$=0 for all other $i$ 
                        & $\beta_{17,j}$=1, $\beta_{42,j}$=3, $\beta_{67,j}$=1, $\beta_{88,j}$=1, $\beta_{91,j}$=1, $\beta_{i,j}$=0 for all other $i$
                        & $\beta_{17,j}$=1, $\beta_{42,j}$=3, $\beta_{67,j}$=1, $\beta_{88,j}$=1, $\beta_{91,j}$=1, $\beta_{i,j}$=0 for all other $i$   
                        & $\beta_{17,j}$=1, $\beta_{42,j}$=3, $\beta_{67,j}$=1, $\beta_{88,j}$=1, $\beta_{91,j}$=1, $\beta_{i,j}$=0 for all other $i$   
                        &       N(0,0.1)        &       0\%     \\\cmidrule{4-10}       
                        &    &  & $y_t^2$
                        & $\beta_{17,j}$=1, $\beta_{42,j}$=3, $\beta_{67,j}$=1, $\beta_{88,j}$=1, $\beta_{91,j}$=1, $\beta_{i,j}$=0 for all other $i$ 
                        & $\beta_{17,j}$=1, $\beta_{42,j}$=3, $\beta_{67,j}$=1, $\beta_{88,j}$=1, $\beta_{91,j}$=1, $\beta_{i,j}$=0 for all other $i$
                        & $\beta_{17,j}$=3, $\beta_{42,j}$=3, $\beta_{67,j}$=1, $\beta_{88,j}$=1, $\beta_{91,j}$=1, $\beta_{i,j}$=0 for all other $i$   
                        & $\beta_{17,j}$=3, $\beta_{42,j}$=3, $\beta_{67,j}$=1, $\beta_{88,j}$=1, $\beta_{91,j}$=1, $\beta_{i,j}$=0 for all other $i$   
                        &               &               \\\cmidrule{4-10}       
                        &    &  & $y_t^3$
                        & $\beta_{17,j}$=3, $\beta_{42,j}$=3, $\beta_{67,j}$=1, $\beta_{88,j}$=1, $\beta_{91,j}$=1, $\beta_{i,j}$=0 for all other $i$ 
                        & $\beta_{17,j}$=3, $\beta_{42,j}$=3, $\beta_{67,j}$=1, $\beta_{88,j}$=1, $\beta_{91,j}$=1, $\beta_{i,j}$=0 for all other $i$
                        & $\beta_{17,j}$=3, $\beta_{42,j}$=3, $\beta_{67,j}$=1, $\beta_{88,j}$=1, $\beta_{91,j}$=1, $\beta_{i,j}$=0 for all other $i$   
                        & $\beta_{17,j}$=5, $\beta_{42,j}$=3, $\beta_{67,j}$=1, $\beta_{88,j}$=1, $\beta_{91,j}$=1, $\beta_{i,j}$=0 for all other $i$   
                        &               &               \\\hline\T 
                        
                        (10) &100  & 2 & $y_t^1$
                        & $\beta_{17,j}$=1, $\beta_{42,j}$=1, $\beta_{67,j}$=1, $\beta_{88,j}$=1, $\beta_{91,j}$=1, $\beta_{i,j}$=0 for all other $i$ 
                        & $\beta_{17,j}$=1, $\beta_{42,j}$=3, $\beta_{67,j}$=1, $\beta_{88,j}$=1, $\beta_{91,j}$=1, $\beta_{i,j}$=0 for all other $i$
                        & $\beta_{17,j}$=1, $\beta_{42,j}$=3, $\beta_{67,j}$=1, $\beta_{88,j}$=1, $\beta_{91,j}$=1, $\beta_{i,j}$=0 for all other $i$   
                        & $\beta_{17,j}$=1, $\beta_{42,j}$=3, $\beta_{67,j}$=1, $\beta_{88,j}$=1, $\beta_{91,j}$=1, $\beta_{i,j}$=0 for all other $i$   
                        &       N(0,0.1)        &       10\%    \\\cmidrule{4-10}       
                        &    &  & $y_t^2$
                        & $\beta_{17,j}$=1, $\beta_{42,j}$=3, $\beta_{67,j}$=1, $\beta_{88,j}$=1, $\beta_{91,j}$=1, $\beta_{i,j}$=0 for all other $i$ 
                        & $\beta_{17,j}$=1, $\beta_{42,j}$=3, $\beta_{67,j}$=1, $\beta_{88,j}$=1, $\beta_{91,j}$=1, $\beta_{i,j}$=0 for all other $i$
                        & $\beta_{17,j}$=3, $\beta_{42,j}$=3, $\beta_{67,j}$=1, $\beta_{88,j}$=1, $\beta_{91,j}$=1, $\beta_{i,j}$=0 for all other $i$   
                        & $\beta_{17,j}$=3, $\beta_{42,j}$=3, $\beta_{67,j}$=1, $\beta_{88,j}$=1, $\beta_{91,j}$=1, $\beta_{i,j}$=0 for all other $i$   
                        &               &               \\\cmidrule{4-10}       
                        &    &  & $y_t^3$
                        & $\beta_{17,j}$=3, $\beta_{42,j}$=3, $\beta_{67,j}$=1, $\beta_{88,j}$=1, $\beta_{91,j}$=1, $\beta_{i,j}$=0 for all other $i$ 
                        & $\beta_{17,j}$=3, $\beta_{42,j}$=3, $\beta_{67,j}$=1, $\beta_{88,j}$=1, $\beta_{91,j}$=1, $\beta_{i,j}$=0 for all other $i$
                        & $\beta_{17,j}$=3, $\beta_{42,j}$=3, $\beta_{67,j}$=1, $\beta_{88,j}$=1, $\beta_{91,j}$=1, $\beta_{i,j}$=0 for all other $i$   
                        & $\beta_{17,j}$=5, $\beta_{42,j}$=3, $\beta_{67,j}$=1, $\beta_{88,j}$=1, $\beta_{91,j}$=1, $\beta_{i,j}$=0 for all other $i$   
                        &               &               \Bot \\                                  
       \hline                                                                                                                          
                \end{tabular} 
                }                                                                                                                          
        \end{scriptsize}                                                                                                                                
\end{table}             

As the main benchmark for NSA (Nonparametric
Splitting Algorithm, Section~\ref{sec:NSA}) and DP (Dynamic Programming,
Sections~\ref{sec:known-change}-\ref{sec:unknown-change}), we
consider the most widely adopted structural change detection
algorithm (BP) developed by~\cite{bai-perron98,bai-perron03}. Similar to
our DP, this algorithm uses dynamic programming to find the change points
that are global minimizers of the sum of squared residuals. In BP, the number
of changes is detected by using a sequential method based on a test with
a null hypothesis of $k$ breaks against $k+1$ breaks. In the experiments with multivariate response, we replace BP by its multivariate counterpart proposed by~\cite{qu-perron-07}. Hereafter, known as QP. As a second baseline,
we consider the Parametric Splitting Algorithm (PSA) proposed by~\cite{gorskikh-16},
which is based on parametric assumptions and sequential application of Chow-test. Additionally, we have also included
the ECP\ method by~\cite{matteson-james14} as one of the baselines to be used in the univariate experiments. Though, this method is designed to detect distributional changes in $(y_t, x_t)$ rather than structural changes, it is nevertheless an interesting benchmark. Like our NSA\ and DP algorithms, ECP is also powered by the energy-distance statistics by~\cite{szekely-rizzo05}. Appendix B provides additional information on how these methods were used
in our simulation study. 

To compare the selected algorithms (DP, NSA,
BP/QP, PSA, ECP), we run them over 1000 simulated datasets for all models using the coefficients from Tables~\ref{tab:univariate-coefs} and~\ref{tab:multivariate-coefs}.
Two performance measures are considered. First, we examine the distribution
of $\hat{k}-k$, the difference between number of estimated and  true  change
points, to see how well the algorithms can detect the correct number of change
points. However, this distribution statistic does not take the locations
of the change points into account. Therefore, as a complementary statistic,
we propose 
\begin{equation}\label{R-stat}
R = \sum_{i=1}^{\min(k,\hat{k})}|T_i-\hat{T}_i| + r|k-\hat{k}|,
\end{equation}
which measures the prediction error of an algorithm both in terms of location
as well as number of detected points. Here, $r$ is a penalty calculated for
each model separately as a maximum of the
change location prediction errors among all the methods and model configurations.
The sequences of detected and true change points are denoted by $\hat{T}_1,\dots,\hat{T}_{\hat{k}}$
 and $T_1,...,T_k,$ respectively. Smaller values of the statistic indicate
better performance.

\subsection{Univariate simulation study}

\begin{table}
        \caption{\label{tab:univariate-test}Univariate samples. Distribution
of $\hat{k}-k$ for the various competing methods over 1000 simulated sample
                paths and corresponding R values reflecting prediction accuracy.}
                                                             
        \centering     
        \renewcommand{\arraystretch}{0.5} 
        \fbox{  \begin{tabular}{l c c c c c c c c c} 

                &               &       \multicolumn{7}{c}{$\hat{k}-k$} 
                                                                        
                                             \\      
                \cline{3-9}

                \T

                {\bf Method}    &       {\bf Model}     &       $\mathbf{\leq
-3}$    &       {\bf -2}        &       {\bf -1}        &       {\bf 0} &
      {\bf 1}      &       {\bf 2} &       $\mathbf{\geq 3}$      &     
 R       \\\hline\T      
                NSA     &       (1)     &               &               &
              &       100.0   &               &               &         
     &       7.5     \\      
                DP     &            &               &               &   
           &       100.0   &               &               &            
  &       1.3     \\      
                BP      &            &               &       1.4     &  
    63.8    &       34.8    &               &               &           
   &       25.4    \\      
                PSA     &           &               &               &   
           &       100.0   &               &               &            
  &       0.0   \\
  ECP     &           &       100.0   &               &   
           &               &               &               &            
  &       120.0   \\\T    
                
                NSA     &       (2)     &       0.6     &       7.9     &
      28.4    &       39.5    &       20.5    &       3.1     &         
     &       59.9    \\      
                DP     &            &       0.6     &       7.9     &   
   30.0    &       38.5    &       19.6    &       3.4     &            
  &       48.3    \\      
                BP      &            &       1.0     &       95.1    &  
    3.9     &               &               &               &           
   &       85.3    \\      
                PSA     &           &       99.9    &       0.1     &   
           &               &               &               &            
  &       119.9   \\
   ECP     &           &       100.0   &               &   
           &               &               &               &            
  &       120.0   \\\T   
                NSA     &       (3)     &               &       3.0     &
      39.0    &       57.7    &       0.3     &               &         
     &       29.5    \\      
                DP     &            &               &       2.4     &   
   41.9    &       55.4    &       0.3     &               &            
  &       21.3    \\      
                BP      &           &       1.2     &       97.7    &   
   1.1     &               &               &               &            
  &       82.2    \\      
                PSA     &          &       88.9    &       10.8    &    
  0.3     &               &               &               &             
 &       111.5   \\
  ECP     &           &       100.0   &               &   
           &               &               &               &            
  &       120.0   \\\T   
                NSA     &       (4)     &       17.7    &       40.0    &
      31.0    &       10.1    &       1.1     &       0.1     &         
     &       83.8    \\      
                DP     &            &       17.2    &       40.8    &   
   30.3    &       10.6    &       0.9     &       0.2     &            
  &       71.6    \\      
                BP      &           &       17.5    &       82.4    &   
   0.1     &               &               &               &            
  &       93.3    \\      
                PSA     &           &       100.0   &               &   
           &               &               &               &            
  &       120.0   \\
   ECP     &           &       100.0   &               &   
           &               &               &               &            
  &       120.0   \\\T    
                NSA     &       (5)     &               &               &
              &       100.0   &               &               &         
     &       7.6     \\      
                DP     &            &               &               &   
           &       100.0   &               &               &            
  &       1.1     \\      
                BP      &           &               &               &   
   0.4     &       99.6    &               &               &            
  &       11.3    \\      
                PSA     &           &               &               &   
           &       100.0   &               &               &            
  &       0.0     \\
   ECP     &           &       100.0   &               &   
           &               &               &               &            
  &       120.0   \\\T    
                NSA     &       (6)     &               &       0.3     &
      7.8     &       45.5    &       38.8    &       7.6     &         
     &       46.1    \\      
                DP     &            &               &       0.3     &   
   6.9     &       46.5    &       38.5    &       7.8     &            
  &       31.4    \\      
                BP      &            &               &       2.9     &  
    59.9    &       37.1    &               &               &           
   &       53.1    \\      
                PSA     &           &       99.8    &       0.2     &   
           &               &               &               &            
  &       119.9   \\
   ECP     &           &       100.0   &               &   
           &               &               &               &            
  &       120.0   \\\T    
                NSA     &       (7)     &               &               &
      3.6     &       95.7    &       0.7     &               &         
     &       13.1    \\      
                DP     &            &               &               &   
   4.2     &       95.1    &       0.7     &               &            
  &       8.2     \\      
                BP      &           &               &       6.8     &   
   70.4    &       22.8    &               &               &            
  &       37.5    \\      
                PSA     &           &       88.8    &       10.4    &   
   0.8     &               &               &               &            
  &       111.1   \\
   ECP     &           &       100.0   &               &   
           &               &               &               &            
  &       120.0   \\\T    
                NSA     &       (8)     &       0.9     &       9.8     &
      34.4    &       44.5    &       9.8     &       0.6     &         
     &       46.2    \\      
                DP     &            &       0.7     &       10.4    &   
   33.4    &       44.8    &       9.8     &       0.9     &            
  &       35.1    \\      
                BP      &           &               &       38.8    &   
   57.1    &       4.1     &               &               &            
  &       59.2    \\      
                PSA     &           &       100.0   &               &   
           &               &               &               &            
  &       120.0   \\
   ECP     &           &       100.0   &               &   
           &               &               &               &            
  &       120.0   \\\T    
                NSA     &       (9)     &               &               &
         &       100.0    &          &               &               &  
    9.1    \\      
                DP     &            &               &               &   
       &       n/a    &           &               &               &     
n/a    \\      
                BP      &           &       100.0        &          &   
       &           &               &               &               &    
  120.0    \\      
                PSA     &           &           &           &           &
       100.0       &               &               &               &    
  2.2   \\
   ECP     &           &       100.0   &               &   
           &               &               &               &            
  &       120.0   \\\T    
                NSA     &       (10)     &          &           &       6.4
   &       93.6    &           &            &               &       15.3
   \\      
                DP     &            &           &          &          & 
     n/a   &           &           &               &      n/a    \\     

                BP      &           &     100.0          &           &  
      &           &               &               &               &     
 120.0    \\      
                PSA     &           &      91.3    &      5.7         & 
   3.0          &               &               &               &       
       &       109.8  \\
        ECP     &           &       100.0   &               &   
           &               &               &               &            
  &       120.0   \Bot \\      
                \hline

        \end{tabular}  }

\end{table}    

Results from the univariate simulation study are given in Table~\ref{tab:univariate-test}. Configurations used in models (1) to (10) are found in Table~\ref{tab:univariate-coefs}.
In general, all of the methods (except ECP) are
able to process non-contaminated and normally distributed models
1 and 5 quite well, while the other cases are not so straightforward. The inability of ECP to detect any of the structural change points is largely explained by the fact that the joint distribution of $(y_t, x_t)$ can remain quite similar even though the model coefficients are exposed to structural changes when the explanatory variables follow similar distributions. Change point analysis (in distributional sense) and structural change point analysis are clearly two different problem classes. Hence, in the remaining discussions, we will focus on the comparison of methods that have been specifically designed for structural change point detection (i.e., NSA, DP, BP, PSA).

The experiments highlight the significance of error distribution and presence of outliers on the relative performance of the methods. For
instance, the performance of PSA is
excellent (with virtually 100\% detection rate and $R=0$) for models 1 and
5. However,
problems emerge when some noise is added. In the rest of the cases, PSA fails
to  detect any changes and its R value is largest in each group. At the same
time, the three
other methods (NSA, DP, BP) seem to be more robust. The detection rate observed for the
main benchmark BP is considerably higher than
that for PSA, even though it is heavily influenced by the size of a change
$d$. When small changes are considered (models 1-4), BP tends
to underestimate the number of breaks in these cases. But for larger change
magnitudes (models 5-8), BP becomes a strong competitor. However, the nonparametric
alternatives, DP and NSA, outperform other techniques when configurations
with heavy tailed disturbances or substantial amount of outliers are considered.
Based on the
distribution of $\hat{k}-k,$ both DP and NSA are practically equally good.
However, if $R$ value is taken into account, DP appears to be always
more accurate than
NSA. This is to be expected as DP\ relies on dynamic programming,
while NSA is just a heuristic approximation of the procedure.

One of the main benefits of NSA is its ability to use
regularization techniques to perform variable selection within regimes. When
the number of variables grows  large, also the amount of nuisance variables
is likely to grow proportionately.   To demonstrate the benefits of using
regularization in eliminating nuisance variables, we compare the performance
of the algorithms using models (9) and (10) with 100 explanatory variables, where only 5 variables have non-zero coefficients. Though DP is in general on part with NSA, it failed to solve the problem in the given time. BP, on the other hand, completes on time, but is unable to detect changes due to the substantial amount of noise generated by the nuisance variables. PSA has outstanding performance in the case of model (9) with normally distributed errors and absence of outliers. However, the introduction of outliers in model (10) changes the results in favor of NSA, which appears to be robust against the combined noise produced by nuisance variables and outliers.

\subsection{Multivariate simulation study}

The setup of the multivariate simulation study is relatively similar to the
univariate case, except that we have $q=3$ as the number of response variables.
Otherwise, the test models are configured as described in Table~\ref{tab:multivariate-coefs}.
As our main benchmark, we consider the method introduced by~\cite{qu-perron-07}.
Hereafter, we will refer to it as
QP. It is designed to estimate multiple structural changes that occur at
unknown dates in a system of equations. QP is based on normal errors and
likelihood ratio type statistics.  To evaluate the performance of NSA, DP
and QP, we use the two measures as previously:  the  $R$ statistic defined
in~\eqref{R-stat} and the differences $\hat{k}-k$
between the number of estimated and actual structural change points.

\begin{table}                                                                       
        \caption{\label{multivariatetest} Multivariate samples. Distribution of $\hat{k}-k$ for the various competing methods over 1000 simulated sample paths and corresponding R values reflecting prediction accuracy.}        
                                                                                                                                                                      
                \centering     
                \renewcommand{\arraystretch}{0.5} 
                \fbox{\begin{tabular}{l c c c c c c c c c} 
                        &               &       \multicolumn{7}{c}{$\hat{k}-k$}                                                                                                                        \\      
                        \cline{3-9}                                                                                                                                                             
                        \T                                                                                                                                                              
                        {\bf Method}    &       {\bf Model}     &       $\mathbf{\leq -3}$    &       {\bf -2}        &       {\bf -1}        &       {\bf 0} &       {\bf 1}      &       {\bf 2} &       $\mathbf{\geq 3}$      &       R       \\\hline\T      
                        NSA     &       (1)  &               &               &               &       100.0   &               &               &               &       7.7     \\      
                        DP     &           &               &               &               &       100.0   &               &               &               &       1.3     \\      
                        QP      &           &               &               &       4.0     &       96.0    &               &               &               &       2.1     \\\T    
                        NSA     &       (2)  &               &       9.6     &       27.5    &       40.9    &       17.9    &       3.8     &       0.1     &       64.8    \\      
                        DP     &            &               &       8.6     &       27.9    &       40.0    &       20.2    &       3.1     &       0.1     &       51.4    \\      
                        QP      &           &               &               &       48.0    &       16.0    &       20.0    &       8.0     &       8.0     &       72.7    \\\T    
                        NSA     &       (3)  &               &       19.7    &       49.2    &       28.7    &       2.3     &               &               &       49.1    \\      
                        DP     &           &               &       19.3    &       49.2    &       28.5    &       2.9     &       0.1     &               &       44.7    \\      
                        QP      &           &               &               &       44.0    &       18.0    &       22.0    &       10.0    &       6.0     &       73.6    \\\T    
                        NSA     &       (4)  &               &       25.0    &       40.6    &       25.8    &       7.7     &       0.8     &               &       73.2    \\      
                        DP     &           &               &       25.7    &       39.0    &       27.6    &       6.7     &       1.0     &               &       65.9    \\      
                        QP      &           &               &               &       56.0    &       14.0    &       4.0     &       12.0    &       14.0    &       77.7    \\\T    
                        NSA     &       (5)  &               &               &               &       100.0   &               &               &               &       7.9     \\      
                        DP     &            &               &               &               &       100.0   &               &               &               &       1.0     \\      
                        QP      &           &               &               &               &       100.0   &               &               &               &       1.5     \\\T    
                        NSA     &       (6)  &               &       1.8     &       13.7    &       40.7    &       34.8    &       9.0     &               &       54.9    \\      
                        DP     &           &               &       1.50    &       14.30   &       40.7    &       34.40   &       9.0     &       0.1     &       48.3    \\      
                        QP      &           &               &               &       84.0    &       12.0    &       4.0     &               &               &       69.3    \\\T    
                        NSA     &       (7)  &               &               &       12.7    &       81.6    &       5.7     &               &               &       19.5    \\      
                        DP     &           &               &       0.1     &       12.7    &       81.8    &       5.4     &               &               &       13.9    \\      
                        QP      &           &               &               &       52.0    &       28.0    &       10.0    &       6.0     &       4.0     &       64.1    \\\T    
                        NSA     &       (8)  &               &       9.2     &       32.2    &       38.3    &       18.0    &       2.2     &               &       58.9    \\      
                        DP     &           &               &       9.9     &       32.4    &       38.0    &       17.4    &       2.2     &               &       45.8    \\      
                        QP      &           &               &               &       68.0    &       12.0    &       8.0     &       8.0     &       4.0     &       74.1    \\\T    
                        NSA     &       (9)  &               &               &           &       100.0   &          &               &               &      8.3   \\      
                        DP     &           &               &            &           &       n/a     &            &               &               &       n/a     \\      
                        QP      &           &               &               &          &       n/a     &          &          &          &       n/a    \\\T    
                        NSA     &       (10)  &               &         &       15.5    &       79.3    &       5.5   &           &               &       22.3    \\      
                        DP     &           &               &           &        &       n/a     &           &           &               &       n/a     \\      
                        QP      &           &               &               &          &       n/a     &           &         &           &       n/a   \Bot \\      
                        \hline                                                                                                                                                          
                \end{tabular} }                                                                                                                                                          
                                                                                                                                                  
\end{table}

Table \ref{multivariatetest} shows the results.
Again, all three methods appear to perform well under normality and absence
of noise (models 1 and 5). In these cases, the average detection rate is
almost  100\% and also the prediction errors as measured by $R$ are very
small.
However, adding noise in the form of outliers and/or non-normal error distribution
immediately lowers the accuracy (which appears as higher R scores in the
Table \ref{multivariatetest}), especially when the magnitude of the structural
change is small. For instance, QP recovers only 14\% of the breaks under
model 4, and it has a tendency to underestimate
the number of breaks in all models, except 1 and 5. NSA and DP, on the other
hand, show better
performance in all cases, especially in models 5 to 8. Furthermore, as in
the univariate case, NSA and DP are similar in terms of observed $\hat{k}-k$
differences,
while judging by prediction error measure $R,$ DP always outperforms NSA. The results obtained for the models (9) and (10) with large number of explanatory variables are quite similar to what was observed in the case of univariate response. However, both DP and QP that rely on dynamic programming fail to terminate, while NSA appears to be quite successful in detecting the change points also in the presence of outliers in addition to large number of nuisance variables.

\section{Application to financial news analytics}

Fluctuations in stock prices are commonly attributed to the arrival of public
news. While the continuous flood of news helps investors to stay on top of
important events, it is at the same time increasingly difficult to judge
what is the actual information value of a news item~\citep{koudijs-16,yermack-14,boudoukh-13}.
Considering the large volume of news produced everyday, it is safe to assume
that only a tiny fraction of them will actually be reflected in trading activity.
Moreover, as market efficiency has improved, the lifespan of news has shortened,
which implies that also predictive relations between news and stock prices
are shorter-lived. As a result, statistical models trying to capture these
dependencies will be exposed to structural changes, where both parameter
estimates as well as the set of contributing news variables can vary from
one regime to another in a discontinuous manner. In particular, this is likely
to hold true in times of crisis, which tend to show non-stationary behavior
(M\"unnix et al. 2012).

\subsection{Extraction of events from Reuters news-wire}

To demonstrate our approach in the context of news analytics, we consider
Thomson Reuters financial news-wire data set from years 2006 to 2009, which
covers the recent credit crunch -period that led to the collapse of Lehman
Brothers. While analyzing the data, we are interested in identifying potential
structural breakpoints as well as the subsets of news variables that are
relevant for predicting banking sector returns within corresponding regimes.
The experiment is carried out in two steps: (i) First, we use a  deep neural
network  to annotate news with event tags (see Appendix C.1. for more details). Each tag indicates
whether a certain news-event has been found in a document. (ii) Second, the
event indicators
are then aggregated into time series showing the number of times each event-type
has been mentioned within a given time step. The aggregation is done separately
for each company.  To ensure sufficient news coverage for each bank, the
study was restricted to the following large banks: Bank of America, Bank
of New York Mellon Corp, Citigroup, Capital One Financial Corp, Goldman Sachs,
JP Morgan Chase \& Co, Morgan Stanley, PNC Financial Services Group, U.S.
Bancorp, Wells Fargo \& Co. 

\begin{figure}[!htp]
        \centering
        \includegraphics[width=13cm,height=8cm]{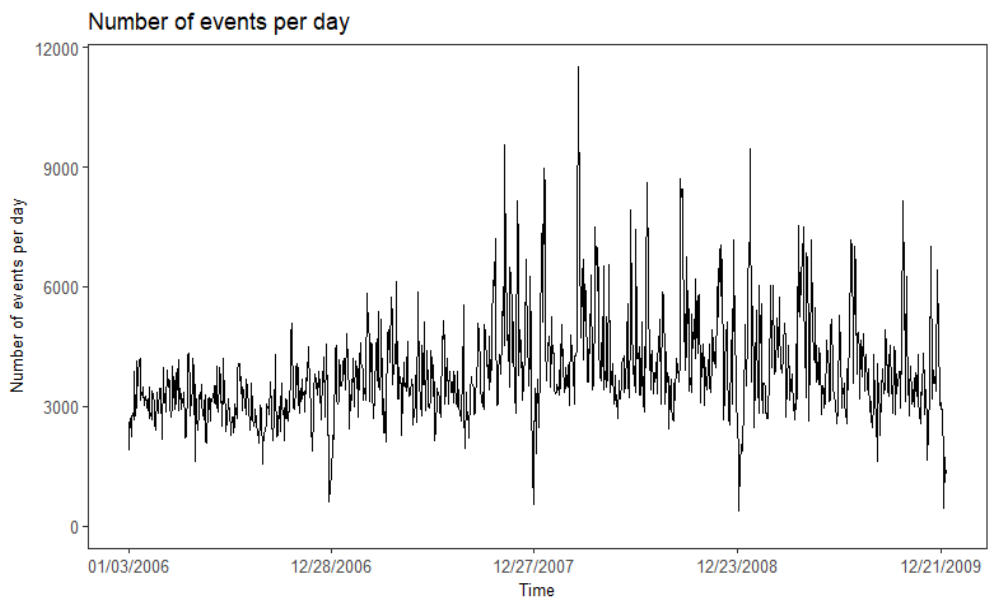}
        \caption{\small Event arrival rates (number of event-type mentions
per day) for major banks.}\label{fig:events-arrival}
\end{figure}

On average, Reuters has published around 530 news per day dealing with
the 10 selected banks. A quick glance at the graph shows that both amount
of news as well as variance in the arrival rate has increased since the beginning
of 2007. The pattern is even more pronounced when considering the number
of events per day as shown in Figure~\ref{fig:events-arrival}. The average
event arrival rate has been around 3850 mentions per day. However, the number
of distinct events is considerably smaller, since there are typically multiple
event-mentions that refer to the same underlying event.  

\subsection{Detection of structural changes during financial crisis}

Next, we applied the non-parametric structural change
detection algorithm NSA on the banking industry returns. The analysis was
done as a multivariate
run covering all banks simultaneously using $L_1$
norm as the regularization function $\varphi$. The regularization strength
parameter $\gamma_n$ was selected using Bayesian information criterion. As a response variable $y_t=(r_t^1, \dots, r_t^{10})\in\reals^{10}$, we consider the log-returns of the 10 banks.
As explanatory variables, we have $x_t=(v_t, e_t)\in\reals^{591}$, where $v_t\in\reals^{10}$ represents the bank-specific trading volumes and $e_t\in\reals^{581}$
is the collection of event count indicators that have been extracted from Reuters news.

\begin{figure}[!h]
        \centering        
        \includegraphics[width=14cm,height=9cm]{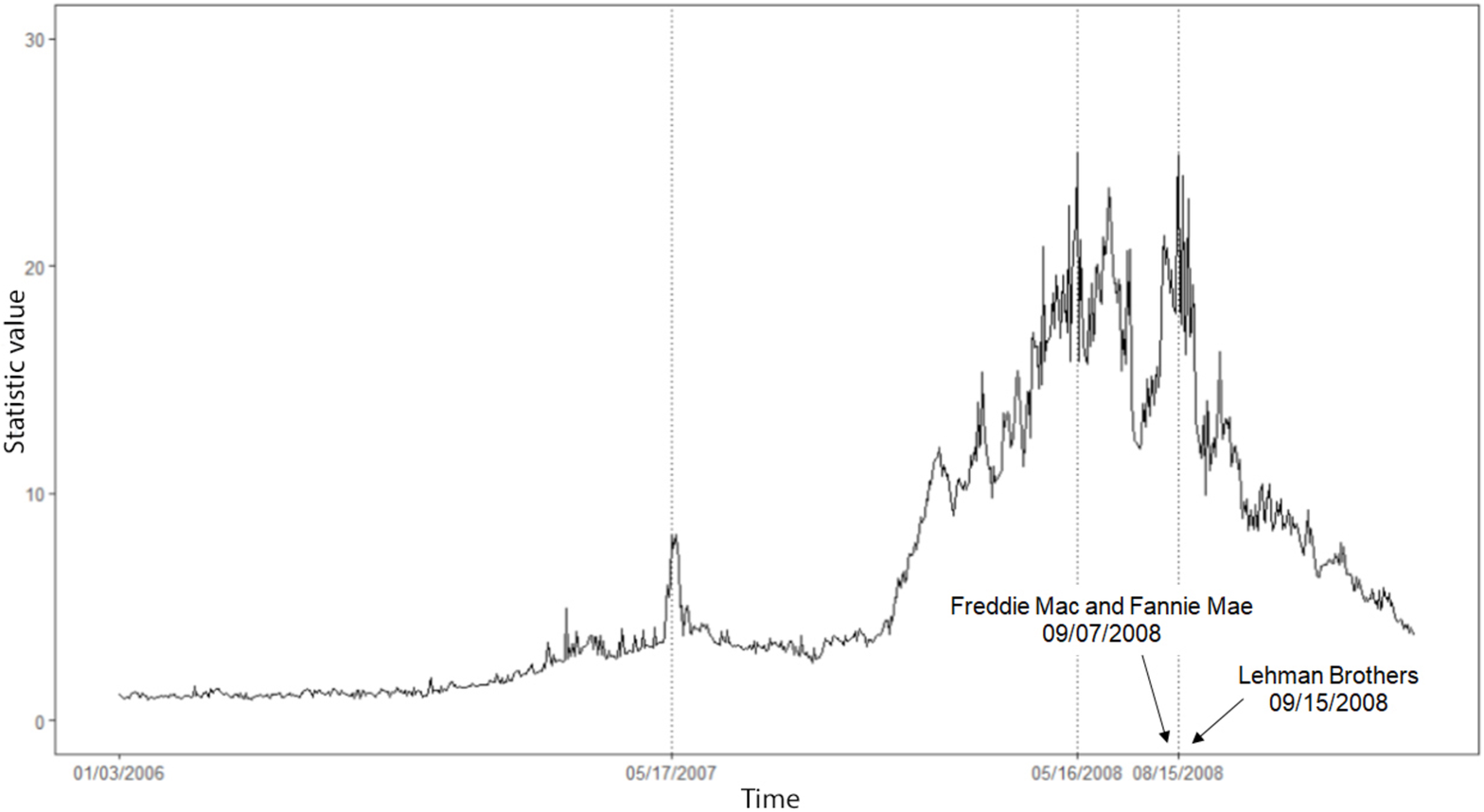}
        
        \caption{\small Multivariate energy-distance statistic. The dashed
vertical lines indicate the locations of structural change points detected
using bootstrap test statistics.}\label{fig:multivariate}
\end{figure}

The
results are shown in Figure~\ref{fig:multivariate}. The multivariate statistic suggests 3 change points, which are located in the middle of May-2007, May-2008 and August-2008. When considering similar statistics for the individual banks, we see a bit more variation in number and location of changes, but they are,
nevertheless, quite close to the ones detected by the multivariate statistic.

\begin{figure}[!h]
        \centering        
         \includegraphics[width=14cm,height=7cm]{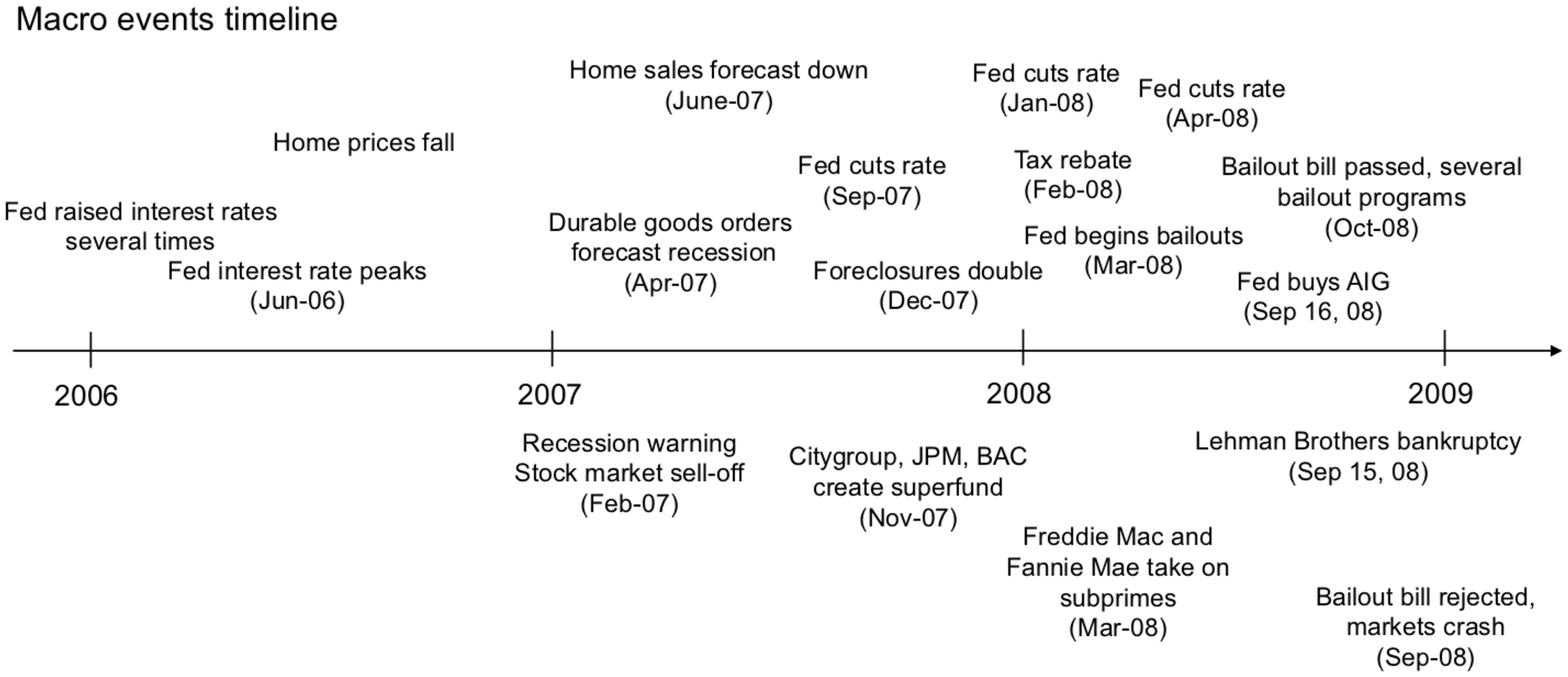}
                \caption{\small Macro-events timeline. }\label{fig:macro-events}
\end{figure}

The macro-events timeline in Figure~\ref{fig:macro-events} gives rather natural
explanations for the four regimes found by the multivariate statistic: (i)
The first regime (01/03/2006 - 05/17/2007) can be interpreted as the escalation
of subprime mortgage bubble into a recession. As home prices fell and Fed
rates remained high, many homeowners couldn't pay their mortgages, nor sell
their homes for a profit. The high number of defaults caused the subprime
mortgage crisis, which by March 2007 was spreading to the financial industry.
(ii) The second regime (05/17-2007 - 05/16/2008) marks the period where the
Fed finally takes action to curb the crisis through sequence of interest
rate cuts and plans for bailout programs. (iii) However, despite the promising
actions, the entire economy was already in recession during the third
regime (05/16/2008 - 08/15/2008). This short and unstable regime soon ended
as the mortgage giants Fannie Mae and Freddie Mac succumbed to the subprime
crisis in August 2008. (iv) Their bankruptcy was soon followed by the cases
of Lehman Brothers and AIG. To prevent the financial system from collapsing,
the fourth regime (08/15/2008 - 01/01/2009) represents the period of massive
bailout programs. For bank-specific analysis, see Appendix C.2.

As a disclaimer applying to this empirical example with financial data, it
is important to note that there are nor 'right' or 'wrong' number of changes.
Here, we have a used rather conservative settings, which allow detection
of only substantial changes in the residual distributions. However, these
settings can be naturally adjusted depending on the use case. For instance,
analysts, who need early warning mechanisms, may  want to use much higher
detection sensitivity. As seen from Figure~\ref{fig:multivariate},
the energy distance statistic shows many spikes that are not considered as
structural changes under the current settings, but which could be really
meaningful as early warning signals that could be utilized by traders and
policy makers alike.  

\section{Conclusions}

We have studied energy-distance based approaches for structural change detection
in linear regression models. In particular, we consider models with multiple
responses and potentially large number of explanatory variables.    Our results
show that already weak moment conditions on regressors and residuals are
sufficient to ensure consistent estimation of structural change points. Furthermore,
our simulation studies show that even under heavy-tailed errors or outlier
contamination, both locations of structural change points and subsets of
contributing variables can still be detected with high accuracy. Two alternative
algorithms are suggested. The first algorithm is based on the use of dynamic
programming principle to find the change points as global minimizers of the
energy-distances between regime-wise residuals. The second algorithm is a
heuristic, which combines nonparametric energy-distance with a computationally
efficient splitting strategy. Though dynamic programming always leads to
better detection accuracy, the heuristic came very close under most test
configurations in the simulation studies. We also demonstrated the importance
of regularization techniques in eliminating nuisance variables from the models
and the subsequent impact on accuracy of structural change detection.

\bibliographystyle{chicago}

\bibliography{highdimensional}

\begin{thebibliography}{}

\bibitem[\protect\citeauthoryear{Abernethy, Bach, Evgeniou, and Vert}{Abernethy
  et~al.}{2009}]{abernethy09}
Abernethy, J., F.~Bach, T.~Evgeniou, and J.-P. Vert (2009).
\newblock A new approach to collaborative filtering: Operator estimation with
  spectral regularization.
\newblock {\em Journal of Machine Learning Research\/}~{\em 10}, 803--826.

\bibitem[\protect\citeauthoryear{Agarwal, Negahban, and Wainwright}{Agarwal
  et~al.}{2012}]{agarwal12}
Agarwal, A., S.~Negahban, and M.~J. Wainwright (2012, 04).
\newblock Noisy matrix decomposition via convex relaxation: Optimal rates in
  high dimensions.
\newblock {\em Ann. Statist.\/}~{\em 40\/}(2), 1171--1197.

\bibitem[\protect\citeauthoryear{Bai and Perron}{Bai and
  Perron}{1998}]{bai-perron98}
Bai, J. and P.~Perron (1998).
\newblock Estimating and testing linear models with multiple structural
  changes.
\newblock {\em Econometrica\/}~{\em 66\/}(1), 47--78.

\bibitem[\protect\citeauthoryear{Bai and Perron}{Bai and
  Perron}{2003}]{bai-perron03}
Bai, J. and P.~Perron (2003).
\newblock Computation and analysis of multiple structural change models.
\newblock {\em Journal of Applied Econometrics\/}~{\em 18\/}(1), 1--22.

\bibitem[\protect\citeauthoryear{Bellman and Roth}{Bellman and
  Roth}{1969}]{bellman-roth69}
Bellman, R. and R.~Roth (1969).
\newblock Curve fitting by segmented straight lines.
\newblock {\em Journal of the American Statistical Association\/}~{\em 64},
  111--125.

\bibitem[\protect\citeauthoryear{Boudoukh, Feldman, Kogan, and
  Richardson}{Boudoukh et~al.}{2013}]{boudoukh-13}
Boudoukh, J., R.~Feldman, S.~Kogan, and M.~Richardson (2013, January).
\newblock Which news moves stock prices? a textual analysis.
\newblock Working Paper 18725, National Bureau of Economic Research.

\bibitem[\protect\citeauthoryear{Cho and Fryzlewicz}{Cho and
  Fryzlewicz}{2015}]{cho-fryzlewicz-15}
Cho, H. and P.~Fryzlewicz (2015).
\newblock Multiple-change-point detection for high dimensional time series via
  sparsified binary segmentation.
\newblock {\em Journal of the Royal Statistical Society: Series B (Statistical
  Methodology)\/}~{\em 77\/}(2), 475--507.

\bibitem[\protect\citeauthoryear{Chopin}{Chopin}{2006}]{chopin06}
Chopin, N. (2006).
\newblock Dynamic detection of change points in long time series.
\newblock {\em The Institute of Statistical Mathematics, pp\/}, 349--366.

\bibitem[\protect\citeauthoryear{Davis, Lee, and Rodriguez-Yam}{Davis
  et~al.}{2006}]{davis-06}
Davis, R., T.~Lee, and G.~Rodriguez-Yam (2006).
\newblock Structural break estimation for nonstationary time series models.
\newblock {\em Journal of the American Statistical Association\/}~{\em
  101\/}(473), 223--239.

\bibitem[\protect\citeauthoryear{Fan, Lv, and Qi}{Fan et~al.}{2011}]{fan11}
Fan, J., J.~Lv, and L.~Qi (2011).
\newblock Sparse high-dimensional models in economics.
\newblock {\em Annual Review of Economics\/}~{\em 3\/}(1), 291--317.

\bibitem[\protect\citeauthoryear{Fisher}{Fisher}{1958}]{fisher58}
Fisher, W.~D. (1958).
\newblock On grouping for maximum homogeneity.
\newblock {\em Journal of the American Statistical Association\/}~{\em 53},
  789--798.

\bibitem[\protect\citeauthoryear{Fryzlewicz}{Fryzlewicz}{2014}]{fryzlewicz-14}
Fryzlewicz, P. (2014, 12).
\newblock Wild binary segmentation for multiple change-point detection.
\newblock {\em Ann. Statist.\/}~{\em 42\/}(6), 2243--2281.

\bibitem[\protect\citeauthoryear{Gorskikh}{Gorskikh}{2016}]{gorskikh-16}
Gorskikh, O. (2016).
\newblock Splitting algorithm for detecting structural changes in predictive
  relationships.
\newblock In {\em Advances in Data Mining. Applications and Theoretical
  Aspects. ICDM 2016. Lecture Notes in Computer Science}, Cham, pp.\  405--419.
  Springer.

\bibitem[\protect\citeauthoryear{Groen, Kapetanios, and Price}{Groen
  et~al.}{2013}]{groen13}
Groen, J., G.~Kapetanios, and S.~Price (2013).
\newblock Multivariate methods for monitoring structural change.
\newblock {\em Journal of Applied Econometrics\/}~{\em 28\/}(2), 250--274.

\bibitem[\protect\citeauthoryear{Harchaoui and L\'evy-Leduc}{Harchaoui and
  L\'evy-Leduc}{2010}]{harchaoui-10}
Harchaoui, Z. and C.~L\'evy-Leduc (2010).
\newblock Multiple change-point estimation with a total variation penalty.
\newblock {\em Journal of the American Statistical Association\/}~{\em
  105\/}(492), 1480--1493.

\bibitem[\protect\citeauthoryear{Hariz, Wylie, and Zhang}{Hariz
  et~al.}{2007}]{hariz-07}
Hariz, S., J.~Wylie, and Q.~Zhang (2007, 08).
\newblock Optimal rate of convergence for nonparametric change-point estimators
  for nonstationary sequences.
\newblock {\em Ann. Statist.\/}~{\em 35\/}(4), 1802--1826.

\bibitem[\protect\citeauthoryear{Hoeffding}{Hoeffding}{1961}]{hoeffding61}
Hoeffding, W. (1961).
\newblock The strong law of large numbers for u-statistics.
\newblock Technical report 302, North Carolina state University.

\bibitem[\protect\citeauthoryear{Kanamori, Hido, and Sugiyama}{Kanamori
  et~al.}{2009}]{kanamori-09}
Kanamori, T., S.~Hido, and M.~Sugiyama (2009, December).
\newblock A least-squares approach to direct importance estimation.
\newblock {\em J. Mach. Learn. Res.\/}~{\em 10}, 1391--1445.

\bibitem[\protect\citeauthoryear{Kawahara and Sugiyama}{Kawahara and
  Sugiyama}{2012}]{kawahara-12}
Kawahara, Y. and M.~Sugiyama (2012, April).
\newblock Sequential change-point detection based on direct density-ratio
  estimation.
\newblock {\em Stat. Anal. Data Min.\/}~{\em 5\/}(2), 114--127.

\bibitem[\protect\citeauthoryear{Koudijs}{Koudijs}{2016}]{koudijs-16}
Koudijs, P. (2016).
\newblock The boats that did not sail: Asset price volatility in a natural
  experiment.
\newblock {\em The Journal of Finance\/}~{\em 71\/}(3), 1185--1226.

\bibitem[\protect\citeauthoryear{Kurozumi and Arai}{Kurozumi and
  Arai}{2007}]{kurozumi-azai-07}
Kurozumi, E. and Y.~Arai (2007).
\newblock Efficient estimation and inference in cointegrating regressions with
  structural change.
\newblock {\em Journal of Time Series Analysis\/}~{\em 28\/}(4), 545--575.

\bibitem[\protect\citeauthoryear{Lavielle and Teyssi{\`e}re}{Lavielle and
  Teyssi{\`e}re}{2006}]{lavielle-06}
Lavielle, M. and G.~Teyssi{\`e}re (2006).
\newblock Detection of multiple change-points in multivariate time series.
\newblock {\em Lithuanian Mathematical Journal\/}~{\em 46\/}(3), 287--306.

\bibitem[\protect\citeauthoryear{Lebarbier}{Lebarbier}{2005}]{lebarbier-05}
Lebarbier, E. (2005).
\newblock Detecting multiple change-points in the mean of gaussian process by
  model selection.
\newblock {\em Signal Processing\/}~{\em 85\/}(4), 717 -- 736.

\bibitem[\protect\citeauthoryear{Li and Perron}{Li and
  Perron}{2017}]{li-perron-17}
Li, Y. and P.~Perron (2017).
\newblock Inference on locally ordered breaks in multiple regressions.
\newblock {\em Econometric Reviews\/}~{\em 36\/}(1-3), 289--353.

\bibitem[\protect\citeauthoryear{Liu, Yamada, Collier, and Sugiyama}{Liu
  et~al.}{2013}]{liu-13}
Liu, S., M.~Yamada, N.~Collier, and M.~Sugiyama (2013).
\newblock Change-point detection in time-series data by relative density-ratio
  estimation.
\newblock {\em Neural Networks\/}~{\em 43}, 72 -- 83.

\bibitem[\protect\citeauthoryear{Matteson and James}{Matteson and
  James}{2014}]{matteson-james14}
Matteson, D. and N.~James (2014).
\newblock A nonparametric approach for multiple change point analysis of
  multivariate data.
\newblock {\em Journal of the American Statistical Association\/}~{\em
  109\/}(505), 334--345.

\bibitem[\protect\citeauthoryear{Negahban and Wainwright}{Negahban and
  Wainwright}{2011}]{negahban-wainwright-2011}
Negahban, S. and M.~Wainwright (2011).
\newblock {Estimation of (Near) Low-Rank Matrices with Noise and
  High-Dimensional Scaling}.
\newblock {\em The Annals of Statistics\/}~{\em 39}, 1069--1097.

\bibitem[\protect\citeauthoryear{Qian and Su}{Qian and Su}{2016}]{qian_su_2016}
Qian, J. and L.~Su (2016).
\newblock Shrinkage estimation of regression models with multiple structural
  changes.
\newblock {\em Econometric Theory\/}~{\em 32\/}(6), 1376--1433.

\bibitem[\protect\citeauthoryear{Qu and Perron}{Qu and
  Perron}{2007}]{qu-perron-07}
Qu, Z. and P.~Perron (2007).
\newblock Estimating and testing structural changes in multivariate
  regressions.
\newblock {\em Econometrica\/}~{\em 75\/}(2), 459--502.

\bibitem[\protect\citeauthoryear{Rizzo and Sz\'ekely}{Rizzo and
  Sz\'ekely}{2010}]{rizzo-szekely10}
Rizzo, M. and G.~J. Sz\'ekely (2010).
\newblock Disco analysis: A nonparametric extension of analysis of variance.
\newblock {\em The Annals of Applied Statistics\/}~{\em 4\/}(2), 1034--1055.

\bibitem[\protect\citeauthoryear{Rizzo and Sz\'ekely}{Rizzo and
  Sz\'ekely}{2016}]{rizzo-szekely16}
Rizzo, M. and G.~J. Sz\'ekely (2016).
\newblock Energy distance.
\newblock {\em WIREs Comput Stat\/}~{\em 8}, 27--38.

\bibitem[\protect\citeauthoryear{Ruggieri and Antonellis}{Ruggieri and
  Antonellis}{2016}]{Ruggieri2016}
Ruggieri, E. and M.~Antonellis (2016).
\newblock An exact approach to bayesian sequential change point detection.
\newblock {\em Computational Statistics \& Data Analysis\/}~{\em 97}, 71 -- 86.

\bibitem[\protect\citeauthoryear{Seo, Kembhavi, Farhadi, and Hajishirzi}{Seo
  et~al.}{2016}]{seo-16}
Seo, M.~J., A.~Kembhavi, A.~Farhadi, and H.~Hajishirzi (2016).
\newblock Bidirectional attention flow for machine comprehension.
\newblock {\em CoRR\/}~{\em abs/1611.01603}.

\bibitem[\protect\citeauthoryear{Stock and Watson}{Stock and
  Watson}{2009}]{stock-watson09}
Stock, J. and M.~Watson (2009).
\newblock {\em Forecasting in Dynamic Factor Models Subject to Structural
  Instability}, pp.\  1--57.
\newblock Oxford University Press.

\bibitem[\protect\citeauthoryear{Sz\'ekely and Rizzo}{Sz\'ekely and
  Rizzo}{2005}]{szekely-rizzo05}
Sz\'ekely, G.~J. and M.~Rizzo (2005).
\newblock Hierarchical clustering via joint between-within distances: Extending
  ward's minimum variance method.
\newblock {\em Journal of Classification\/}~{\em 22\/}(2), 151--183.

\bibitem[\protect\citeauthoryear{Sz\'ekely and Rizzo}{Sz\'ekely and
  Rizzo}{2014a}]{szekely-rizzo13}
Sz\'ekely, G.~J. and M.~Rizzo (2014a).
\newblock Energy statistics: A class of statistics based on distances.
\newblock {\em J. Statist. Plann. Inference\/}~{\em 143}, 1249--1272.

\bibitem[\protect\citeauthoryear{Sz\'ekely and Rizzo}{Sz\'ekely and
  Rizzo}{2014b}]{szekely-rizzo14}
Sz\'ekely, G.~J. and M.~Rizzo (2014b).
\newblock Partial distance correlation with methods for dissimilarities.
\newblock {\em The Annals of Statistics\/}~{\em 42\/}(6), 2382--2412.

\bibitem[\protect\citeauthoryear{Yang, Yang, Dyer, He, Smola, and Hovy}{Yang
  et~al.}{2016}]{yang-16}
Yang, Z., D.~Yang, C.~Dyer, X.~He, A.~Smola, and E.~Hovy (2016).
\newblock Hierarchical attention networks for document classification.
\newblock In {\em Proceedings of the 2016 Conference of the North American
  Chapter of the Association for Computational Linguistics: Human Language
  Technologies}, pp.\  1480--1489. Association for Computational Linguistics.

\bibitem[\protect\citeauthoryear{Yermack}{Yermack}{2014}]{yermack-14}
Yermack, D. (2014).
\newblock Tailspotting: Identifying and profiting from ceo vacation trips.
\newblock {\em Journal of Financial Economics\/}~{\em 113\/}(2), 252 -- 269.

\bibitem[\protect\citeauthoryear{Yuan, Ekici, Lu, and Monteiro}{Yuan
  et~al.}{2007}]{yuan07}
Yuan, M., A.~Ekici, Z.~Lu, and R.~Monteiro (2007).
\newblock Dimension reduction and coefficient estimation in multivariate linear
  regression.
\newblock {\em Journal of the Royal Statistical Society: Series B (Statistical
  Methodology)\/}~{\em 69\/}(3), 329--346.

\bibitem[\protect\citeauthoryear{Zeileis}{Zeileis}{2003}]{Zeileis2003}
Zeileis, A., K. C. K. W. H.~K. (2003).
\newblock Testing and dating of structural changes in practice.
\newblock {\em Computational Statistics \& Data Analysis\/}~{\em 44}, 109 --
  123.

\bibitem[\protect\citeauthoryear{Zeileis}{Zeileis}{2001}]{Zeileis2001}
Zeileis, A., L. F. H. K. K.~C. (2001).
\newblock {strucchange. An R package for testing for structural change in
  linear regression models}.
\newblock {Technical report. SFB Adaptive Information Systems and Modelling in
  Economics and Management Science.}, WU Vienna University of Economics and
  Business.

\end{thebibliography}

\section*{Appendix A: Technical proofs for consistency results}
\subsection*{A.1. Proof (of Lemma 2)}
Since
$$
\frac{1}{n^2}\sum_{i_1,i_2=1}^n X_{i_1,i_2} -  C = \frac{1}{n^2}\sum_{i_1,i_2=1}^n
(X_{i_1,i_2} -  C), 
$$
we can, without loss of generality, assume that $C=0$. Minkowski inequality
implies that, for any $a>0$,
$$
\mathbb{P}\left(\frac{1}{n^2}\left|\sum_{i_1,i_2=1}^n X_{i_1,i_2}\right|
> a\right) \leq \frac{1}{n^4a^2}\E\left(\sum_{i_1,i_2=1}^n X_{i_1,i_2}\right)^2=\frac{1}{n^4a^2}\sum_{i_1,i_2,i_3,i_4=1}^n
\E\left[X_{i_1,i_2}X_{i_3,i_4}\right].
$$
Denote
$$
m = \min_{k,j\in\{1,2,3,4\},k\neq j}|i_k-i_j|.
$$
By assumption, for each $\epsilon>0$, there exists $N$ such that $|\E\left[X_{i_1,i_2}X_{i_3,i_4}\right]|<\epsilon$
for any $m\geq N$. Write
$$
\sum_{i_1,i_2,i_3,i_4=1}^n \E\left[X_{i_1,i_2}X_{i_3,i_4}\right] = \sum_{i_1,i_2,i_3,i_4=1,
	m < N}^n \E\left[X_{i_1,i_2}X_{i_3,i_4}\right] + \sum_{i_1,i_2,i_3,i_4=1,
	m\geq N}^n \E\left[X_{i_1,i_2}X_{i_3,i_4}\right].
$$
Since $m<N$ only if the distance between one of the pairs $(i_k,i_l)$ is
less than $N$, we observe that the first term is bounded by 
$$
\sum_{i_1,i_2,i_3,i_4=1, m < N}^n \E\left[X_{i_1,i_2}X_{i_3,i_4}\right] \leq
D(N)n^3
$$
for some finite constant depending only on $N$. For the second term, we estimate
$$
\sum_{i_1,i_2,i_3,i_4=1, m\geq N}^n |\E\left[X_{i_1,i_2}X_{i_3,i_4}\right]|
< \epsilon n^4.
$$
Combining the above bounds we obtain
$$
\mathbb{P}\left(\frac{1}{n^2}\left|\sum_{i_1,i_2=1}^n X_{i_1,i_2}\right|
> a\right) \leq \frac{D(N)}{na^2} + \frac{\epsilon}{a^2}.
$$
Since $\epsilon>0$ is arbitrary, the result follows by choosing $n$ large
enough.

\begin{remark}
	Note that, if 
	\begin{equation}
	\label{eq:doublearray_exp}
	n^{-2}\sum_{i_1,i_2=1}^n \E[X_{i_1,i_2}] \rightarrow C,
	\end{equation}
	then a slight modification of the above proof shows that $\frac{1}{n^2}\sum_{i_1,i_2=1}^n
	X_{i_1,i_2} \rightarrow C$ in probability, provided that $
	Cov(X_{i_1,i_2},X_{i_3,i_4}) \rightarrow 0$ as $\min_{k,j\in\{1,2,3,4\},k\neq
		j}|i_k-i_j| \rightarrow \infty.$ Finally, we note that with similar arguments
	we obtain \eqref{eq:doublearray_exp} provided that $\lim_{|i_1-i_2|\rightarrow
		\infty} |\E[X_{i_1,i_2}]-C| = 0.$
\end{remark}

\subsection*{A.2. Proof (of Proposition 1)}

The proof is by contradiction. Assume that $\lambda_1^0$ is not consistently
estimated, i.e. $\hat{\lambda}_1 \neq \lambda_1^0$. Without loss of generality,
we assume that the estimated change
point $\hat{T}_1$ satisfies $\hat{T}_1 < T\lambda_1^0$, giving
us a partitioning $I_1=[0, \hat{T}_1]$, $I_2=[\hat{T}_1+1, T_1^0]$ and $I_3=[T_1^0+1,T]$.

We denote by $|A|$ the size of a set $A$. In particular, we have that $|I_1
\cup
I_2 \cup I_3| = T$, $|I_3|=(1-\lambda_1^0)T$, $|I_1| \sim \hat{\lambda}_1T$,
$|I_2|\sim(\lambda_1^0 -\hat{\lambda}_1)T$, and $|I_2 \cup I_3| \sim (1-\hat{\lambda}_1)T$
(where the notation $f(T) \sim g(T)$ refers to the usual interpretation $\lim_{T\to\infty}
\frac{f(T)}{g(T)} = 1$). For
notational simplicity, we denote by $\hat{\beta}(i)$ the estimator corresponding
to the region where $i$ belongs. That is, for $\hat{\beta}_i,i=1,2$ denoting
the regularized estimates, we have $\hat\beta(i) = \hat{\beta}_1$ for all
$i \in I_1$, and 
$\hat{\beta}(i) = \hat{\beta}_2$ for all $i \in I_2 \cup I_3$. Similarly,
we denote by $\beta^0(i)$ the correct value corresponding to the region where
$i$ belongs. That is, as the true change point is $T_1^0$, we have $\beta^0(i)
= \beta^0_1$ for all $i \in I_1 \cup I_2$ and $\beta^0(i) = \beta^0_2$ for
all $i\in I_3$. We also denote by $\beta(i)$ and $\beta^c_k$ the limits related
to Assumption 4. More precisely, we always have $\hat\beta(i)
\rightarrow \beta(i)$ and $\hat\beta_k \rightarrow \beta^c_k$. Moreover,
we have  
$\hat\beta(i) = \hat\beta_1\rightarrow \beta^c_1=\beta^0_1$ for all $i \in
I_1$, as region $I_1$ is a subset of the correct interval $[0,T_1^0]$. For
$i \in I_2 \cup I_3$, we have $\hat{\beta}(i) = \hat{\beta}_2 \rightarrow
\beta^c_2$, and thus $\beta(i) = \beta^c_2$.  

Denote by $\hat{u}_i = u_i - x_i'(\hat{\beta}(i)-\beta^0(i)),i\in I_k,k=1,2,3$
the corresponding estimated residuals and let $U_1 = \{\hat{u}_t\}_{t\in
	I_1}$ and $U_2 = \{\hat{u}_t\}_{t\in
	I_2\cup I_3}$ denote the collections of regularized residuals from different
intervals. We set
\[
\hat{\mu}^{\alpha}_{U_1,U_2}  =\frac{1}{|I_1||I_2\cup I_3|}\sum_{j \in I_1,
	i \in I_2\cup I_3} |\hat{u}_i-\hat{u}_j|^{\alpha}, \hat{\mu}^{\alpha}_{U_1}
= \frac{1}{|I_1|^2}\sum_{i,j \in I_1}|\hat{u}_i-\hat{u}_j|^{\alpha},
\]
and
\[
\hat{\mu}^{\alpha}_{U_2}  = \frac{1}{|I_2\cup I_3|^2}\sum_{i,j\in I_2\cup
	I_3}|\hat{u}_i-\hat{u}_j|^{\alpha}.
\]
We prove that 
\begin{equation}
\label{eq:empirical_limit}
2\hat{\mu}^{\alpha}_{U_1,U_2}-\hat{\mu}^{\alpha}_{U_1}-\hat{\mu}^{\alpha}_{U_2}
\rightarrow C>0,
\end{equation}
where $C$ is a constant and the convergence holds in probability. From this
we get
$$
d_{\alpha}(U_1, U_2) = \frac{|I_1||I_2\cup I_3|}{2|I_1|+2|I_2\cup I_3|}\left(2\hat{\mu}^{\alpha}_{U_1,U_2}-\hat{\mu}^{\alpha}_{U_1}-\hat{\mu}^{\alpha}_{U_2}\right)
\rightarrow \infty.
$$
Consequently, $\hat{T}_1$ cannot be a minimizer for  the model equation 2 of Section 2 (in the paper),
which leads to the expected contradiction. 

We divide the rest of the proof into three steps. In step 1 we consider the
differences $\hat{u}_i - \hat{u}_j$ that depend on the entire data set. In
step 2 we calculate the limits of the terms $\hat{\mu}^{\alpha}_{U_1,U_2}$,
$\hat{\mu}^{\alpha}_{U_1}$, and $\hat{\mu}^{\alpha}_{U_2}$. Finally, in step
3, we show \eqref{eq:empirical_limit}.

\textbf{Step 1: We show that, for any subsets $A,B \subset \{1,2,\ldots,T\}$
	we have
	\begin{equation*}
	\begin{split}
	&\lim_{T\rightarrow \infty}\frac{1}{|A||B|}\sum_{i\in A,j \in B}|\hat{u}_i
	- \hat{u}_j|^\alpha \\ 
	&= \lim_{T\rightarrow \infty}\frac{1}{|A||B|}\sum_{i\in A,j \in B}|u_i -
	u_j -x_i'(\beta(i)-\beta^0(i)) + x_j'(\beta(j)-\beta^0(j))|^\alpha,
	\end{split}
	\end{equation*}
	where the limits are understood in probability.}\\
Recall that $\hat{u}_i = u_i - x_i'(\hat{\beta}(i)-\beta^0(i))$ and denote

\begin{equation}
\label{eq:a_ij}
a_{ij} = \hat{u}_i - \hat{u}_j = u_i - x_i'(\hat{\beta}(i)-\beta^0(i)) -
u_j + x_j'(\hat{\beta}(j)-\beta^0(j)),
\end{equation}
and
\begin{equation}
\label{eq:b_ij}
b_{ij} = u_i - x_i'(\beta(i)-\beta^0(i)) - u_j + x_j'(\beta(j)-\beta^0(j)).
\end{equation}
By writing
$$
|a_{ij}|^\alpha = |b_{ij}|^\alpha + |a_{ij}|^{\alpha} - |b_{ij}|^\alpha
$$
it suffices to prove that 
$$
\lim_{T\rightarrow \infty}\frac{1}{|A||B|}\sum_{i\in A,j \in B}||a_{ij}|^{\alpha}
- |b_{ij}|^\alpha| \rightarrow 0
$$
in probability. We now treat the case $\alpha \in(0,1]$ and $\alpha\in(1,2)$
separately. 

\textbf{Step 1.1: $\alpha\in(0,1]$.} \\ 
By using the inequality $
||a|^\alpha - |b|^\alpha| \leq ||a|-|b||^\alpha \leq |a-b|^\alpha
$, valid for all $a,b\in \R^q$ and $\alpha\in(0,1]$,
for $a_{ij}$ and $b_{ij}$ we observe
$$
||a_{ij}|^\alpha - |b_{ij}|^\alpha|\leq |a_{ij}-b_{ij}|^\alpha.
$$
Here, by using $(|a|+|b|)^\alpha \leq |a|^\alpha + |b|^\alpha$, we obtain

\begin{equation*}
\begin{split}
|a_{ij}-b_{ij}|^\alpha &= |x'_i(\beta(i)-\hat{\beta}(i)) + x'_j(\hat{\beta}(j)
- \beta(j))|^\alpha \\
&\leq |x'_i(\hat{\beta}(i))-\beta(i))|^\alpha + |x'_j(\hat{\beta}(j) - \beta(j))|^\alpha
\\
&\leq |x_i|^\alpha \quad ||\hat{\beta}(i))-\beta(i)||^\alpha + |x_j|^\alpha
\quad ||\hat{\beta}(j))-\beta(j)||^\alpha.
\end{split}
\end{equation*}
Since $\hat{\beta}(i)-\beta(i) = \hat{\beta}_1 - \beta^c_1$ for $i \in I_1$
and $\hat{\beta}(i)-\beta(i) = \hat{\beta}_2 - \beta^c_2$ for $i \in I_2\cup
I_3$, we have
\begin{equation*}
\begin{split}
&\lim_{T\rightarrow \infty}\frac{1}{|A||B|}\sum_{i\in A,j \in B}|x_i|^\alpha
\quad ||\hat{\beta}(i))-\beta(i)||^\alpha \\
&= \lim_{T\rightarrow \infty}\frac{1}{|A|}\sum_{i\in A}|x_i|^\alpha \quad
||\hat{\beta}(i))-\beta(i)||^\alpha\\
&=\lim_{T\rightarrow \infty}\frac{1}{|A|}\sum_{k=1}^3\sum_{i\in A \cap I_k}|x_i|^\alpha
\quad ||\hat{\beta}(i))-\beta(i)||^\alpha \\
&\leq \lim_{T\rightarrow \infty}\max_{k\in\{1,2\}}||\hat{\beta}_k-\beta^c_k||^\alpha\frac{1}{|A|}\sum_{i\in
	A}|x_i|^\alpha.
\end{split}
\end{equation*}
Here the random variable 
$$
\frac{1}{|A|}\sum_{i\in A}|x_i|^\alpha
$$ 
is uniformly bounded in $L^1$, and hence also in probability. Furthermore,
we have
$$
\max_{k\in\{1,2\}}||\hat{\beta}_k-\beta^c_k||^\alpha\rightarrow 0
$$
in probability, and thus
$$
\lim_{T\rightarrow \infty}\frac{1}{|A||B|}\sum_{i\in A,j \in B}|x_i|^\alpha
\quad ||\hat{\beta}(i))-\beta(i)||^\alpha \rightarrow 0.
$$
Treating the term $|x_j|^\alpha \quad ||\hat{\beta}(j))-\beta(j)||^\alpha$
similarly yields the claim.\\
\textbf{Step 1.2: $\alpha\in(1,2)$.} \\
We use the following inequality, valid for all $p\geq 0$ and $a,b\in \R^q$:
$$
||a|^p - |b|^p| \leq \max(p,1)2^{(p-2)_+}\left[|a-b|^p + |b|^{(p-1)_+}|a-b|^{\min(p,1)}\right].
$$
Plugging $p=\alpha$ together with $a_{ij}$ and $b_{ij}$ defined in equations
\eqref{eq:a_ij} and \eqref{eq:b_ij} we have
\begin{equation}
\label{eq:a_ij-b_ij}
||a_{ij}|^\alpha - |b_{ij}|^\alpha| \leq \alpha\left[|a_{ij}-b_{ij}|^\alpha
+ |b_{ij}|^{\alpha-1}|a_{ij}-b_{ij}|\right].
\end{equation}
Jensen inequality implies 
\begin{equation*}
\begin{split}
|a_{ij}-b_{ij}|^\alpha =& |x'_i(\beta(i)-\hat{\beta}(i)) + x'_j(\hat{\beta}(j)
- \beta(j))|^\alpha \\
\leq & 2^{\alpha-1}\left[|x_i|^\alpha \quad ||\hat{\beta}(i))-\beta(i)||^\alpha
+ |x_j|^\alpha \quad ||\hat{\beta}(j))-\beta(j)||^\alpha\right].
\end{split}
\end{equation*}
Now the first term on the right-hand side of \eqref{eq:a_ij-b_ij} can be
treated as in the case $\alpha\in(0,1)$. For the second term on the right-hand
side of \eqref{eq:a_ij-b_ij}, 
we apply inequality $|a+b|^{\alpha-1} \leq |a|^{\alpha-1} + |b|^{\alpha-1}$
to estimate
\begin{equation*}
\begin{split}
|b_{ij}|^{\alpha-1} &= |u_i - x_i'(\beta(i)-\beta^0(i)) - u_j + x_j'(\beta(j)-\beta^0(j))|^{\alpha-1}\\
&\leq |u_i|^{\alpha-1} + |x_i|^{\alpha-1} \quad ||\beta(i))-\beta^0(i)||^{\alpha-1}
+ |u_j|^{\alpha-1} + |x_j|^{\alpha-1} \quad ||\beta(j))-\beta^0(j)||^{\alpha-1}.
\end{split}
\end{equation*}
Using this together with
$$
|a_{ij}-b_{ij}| \leq |x_i| \quad ||\hat{\beta}(i))-\beta(i)|| + |x_j|
\quad ||\hat{\beta}(j))-\beta(j)||
$$
and the fact that $\E|u|^{\alpha-1} \leq \left[\E|u|^{\alpha}\right]^{\frac{\alpha-1}{\alpha}}$,
we obtain the claim by following similar steps as in the case $\alpha\in(0,1]$.

\textbf{Step 2: We show that, for the limit $L$ defined by
	$$
	L:=\lim_{T\rightarrow \infty}\left(2\hat{\mu}^{\alpha}_{U_1,U_2}-\hat{\mu}^{\alpha}_{U_1}-\hat{\mu}^{\alpha}_{U_2}\right),
	$$
	we have 
	\begin{equation*}
	\begin{split}
	L&= 
	\frac{2(\lambda_1^0 - \hat{\lambda}_1)}{1- \hat{\lambda}_1}\E|U-\tilde{U}+X_1'(\beta^c_2
	- \beta^0_1)|^\alpha \\
	&+\frac{2(1 - \lambda_1^0)}{1-\hat\lambda_1}\E|U-\tilde{U}-X_2'(\beta^c_2
	- \beta^0_2)|^\alpha - \E|U-\tilde{U}|^\alpha \\
	&-\frac{(\lambda_1^0 - \hat\lambda_1)^2}{(1-\hat\lambda_1)^2}\E|U - \tilde{U}
	+ (\tilde{X}_1'-X_1')(\beta^c_2-\beta^0_1)|^\alpha \\
	&-\frac{(1 - \lambda_1^0)^2}{(1-\hat\lambda_1)^2}\E|U - \tilde{U} + (\tilde{X}_2'-X_2')(\beta^c_2-\beta^0_2)|^\alpha
	\\
	&- 2 \frac{(1 - \lambda_1^0)(\lambda_1^0 - \hat\lambda_1)}{(1-\hat\lambda_1)^2}\E|U
	- \tilde{U} -X_2'(\beta^c_2-\beta^0_2) + \tilde{X}_1'(\beta^c_2-\beta^0_1)|^\alpha,
	\end{split}
	\end{equation*}
	where $U,\tilde{U}$ are independent copies of disturbances $u$ and $X_i',\tilde{X}_i'$
	are independent copies drawn from the distribution of $X_i$ given in Assumption
	A2.}\\
We study the limits of the terms $\hat{\mu}^{\alpha}_{U_1,U_2}$, $\hat{\mu}^{\alpha}_{U_1}$,
and $\hat{\mu}^{\alpha}_{U_2}$ separately. For the term $\hat{\mu}^{\alpha}_{U_1}$,
as $\beta(i) = \beta^0(i)$ for all $i\in I_1$, we observe
$$
\hat{\mu}^{\alpha}_{U_1} \rightarrow \E|U-\tilde{U}|^\alpha.
$$
Consider next the term 
$\hat{\mu}^{\alpha}_{U_1,U_2}$. Since $\beta(i) = \beta^0(i)$ for all $i
\in I_1$, step 1 implies that it suffices to study the limit
\begin{equation*}
\begin{split}
&\lim_{T\rightarrow \infty}\frac{1}{|I_1||I_2\cup I_3|}\sum_{i\in I_1,j\in
	I_2\cup I_3}|u_i - u_j + x_j'(\beta(j)-\beta^0(j))|^\alpha \\
=& \lim_{T\rightarrow \infty}\frac{1}{|I_1||I_2\cup I_3|}\sum_{i\in I_1,j\in
	I_2}|u_i - u_j + x_j'(\beta(j)-\beta^0(j))|^\alpha \\
&+ \lim_{T\rightarrow \infty}\frac{1}{|I_1||I_2\cup I_3|}\sum_{i\in I_1,j\in
	I_3}|u_i - u_j + x_j'(\beta(j)-\beta^0(j))|^\alpha.   
\end{split}
\end{equation*}
Recall that $\beta(j)=\beta^c_2$, $\beta^0(j)=\beta_1^0$, and $x_j \sim X_1$
for all $j\in I_2$. Since the proportion of observations is $\lambda_1^0
- \hat\lambda_1$ in the regime $I_2$ and $1-\hat\lambda_1$ in the regime
$I_2 \cup I_3$, it now follows from Assumption 5 and Lemma
2, 
that
$$
\lim_{T\rightarrow \infty}\frac{1}{|I_1||I_2\cup I_3|}\sum_{i\in I_1,j\in
	I_2}|u_i - u_j + x_j'(\beta(j)-\beta^0(j))|^\alpha = \frac{\lambda_1^0 -
	\hat{\lambda}_1}{1- \hat{\lambda}_1}\E|U-\tilde{U}+X_1'(\beta^c_2 - \beta^0_1)|^\alpha.
$$
Similarly, as $\beta(j) = \beta^c_2$, $\beta^0(j) = \beta^0_2$, and $x_j
\sim X_2$ for all $j\in I_3$, we observe that
$$
\lim_{T\rightarrow \infty}\frac{1}{|I_1||I_2\cup I_3|}\sum_{i\in I_1,j\in
	I_3}|u_i - u_j + x_j'(\beta(j)-\beta^0(j))|^\alpha = \frac{1 - \lambda_1^0}{1-\hat\lambda_1}\E|U-\tilde{U}-X_2'(\beta^c_2
- \beta^0_2)|^\alpha.
$$
It remains to study the limit of the term $\hat{\mu}^{\alpha}_{U_2}$. As
above, we split 
$$
\{i \in I_2 \cup I_3, j\in I_2\cup I_3\} = \{i,j \in I_2\} \cup \{i,j\in
I_3\} \cup \{i\in I_2, j\in I_3\} \cup \{i \in I_3, j \in I_2\}.
$$
Similarly as above, we can apply Assumption 5 and Lemma
2 to obtain
\begin{equation*}
\begin{split}
&\lim_{T\rightarrow \infty}\frac{1}{|I_2\cup I_3|^2}\sum_{i,j\in I_2}|u_i
- u_j -x_i'(\beta(i)-\beta^0(i)) + x_j'(\beta(j)-\beta^0(j))|^\alpha \\
=& \frac{(\lambda_1^0 - \hat\lambda_1)^2}{(1-\hat\lambda_1)^2}\E|U-\tilde{U}
-X_1'(\beta^c_2-\beta^0_1) + \tilde{X}_1'(\beta^c_2-\beta^0_1)|^\alpha,
\end{split}
\end{equation*}
\begin{equation*}
\begin{split}
&\lim_{T\rightarrow \infty}\frac{1}{|I_2\cup I_3|^2}\sum_{i,j\in I_3}|u_i
- u_j -x_i'(\beta(i)-\beta^0(i)) + x_j'(\beta(j)-\beta^0(j))|^\alpha \\
=& \frac{(1 - \lambda_1^0)^2}{(1-\hat\lambda_1)^2}\E|U-\tilde{U} -X_2'(\beta^c_2-\beta^0_2)
+ \tilde{X}_2'(\beta^c_2-\beta^0_2)|^\alpha,
\end{split}
\end{equation*}
\begin{equation*}
\begin{split}
&\lim_{T\rightarrow \infty}\frac{1}{|I_2\cup I_3|^2}\sum_{i\in I_2,j\in I_3}|u_i
- u_j -x_i'(\beta(i)-\beta^0(i)) + x_j'(\beta(j)-\beta^0(j))|^\alpha \\
=& \frac{(1 - \lambda_1^0)(\lambda_1^0 - \hat\lambda_1)}{(1-\hat\lambda_1)^2}\E|U-\tilde{U}
-X_1'(\beta^c_2-\beta^0_1) + \tilde{X}_2'(\beta^c_2-\beta^0_2)|^\alpha,
\end{split}
\end{equation*}
and
\begin{equation*}
\begin{split}
&\lim_{T\rightarrow \infty}\frac{1}{|I_2\cup I_3|^2}\sum_{i\in I_3,j\in I_2}|u_i
- u_j -x_i'(\beta(i)-\beta^0(i)) + x_j'(\beta(j)-\beta^0(j))|^\alpha \\
=& \frac{(1 - \lambda_1^0)(\lambda_1^0 - \hat\lambda_1)}{(1-\hat\lambda_1)^2}\E|U-\tilde{U}
-X_2'(\beta^c_2-\beta^0_2) + \tilde{X}_1'(\beta^c_2-\beta^0_1)|^\alpha.
\end{split}
\end{equation*}
Observing
\begin{equation*}
\E|U-\tilde{U} -X_1'(\beta^c_2-\beta^0_1) + \tilde{X}_2'(\beta^c_2-\beta^0_2)|^\alpha
= \E|U-\tilde{U} -X_2'(\beta^c_2-\beta^0_2) + \tilde{X}_1'(\beta^c_2-\beta^0_1)|^\alpha
\end{equation*}
we obtain the claim.

\textbf{Step 3: We show that for the limit $L$ defined in step 2, we have
	$L>0$.}\\
By definition of the energy distance $\mathcal{E}(u,v;\alpha)$ for two random
variables $u$ and $v$, we have that
\begin{equation*}
\begin{split}
\mathcal{E}(U,U+X_1'(\beta^c_2-\beta^0_1);\alpha) 
&= 2\E|U-\tilde{U}+X_1'(\beta^c_2 - \beta^0_1)|^\alpha - \E|U-\tilde{U}|^\alpha
\\
&- \E|U-\tilde{U} + (\tilde{X}_1'-X_1')(\beta^c_2-\beta^0_1)|^\alpha,
\end{split}
\end{equation*}
\begin{equation*}
\begin{split}
\mathcal{E}(U,U+X_2'(\beta^c_2-\beta^0_2);\alpha) &= 2\E|U-\tilde{U}+X_2'(\beta^c_2
- \beta^0_2)|^\alpha - \E|U-\tilde{U}|^\alpha \\
&- \E|U-\tilde{U} + (\tilde{X}_2'-X_2')(\beta^c_2-\beta^0_2)|^\alpha,
\end{split}
\end{equation*}
and
\begin{equation*}
\begin{split}
\mathcal{E}(U+X_2'(\beta^c_2-\beta^0_2),U+X_1'(\beta^c_2-\beta^0_1);\alpha)&=2\E|U-\tilde{U}+X_1'(\beta^c_2-\beta^0_1)-\tilde{X}_2'(\beta^c_2-\beta^0_2)|^\alpha
\\
&- \E|U-\tilde{U}+(X_1'-\tilde{X}_1')(\beta^c_2-\beta^0_1)|^\alpha\\
& - \E|U-\tilde{U}+(X_2'-\tilde{X}_2')(\beta^c_2-\beta^0_2)|^\alpha.
\end{split}
\end{equation*}
Moreover, we have that
\begin{equation*}
\E|U-\tilde{U}|^\alpha = \frac{\lambda_1^0 - \hat{\lambda}_1}{1- \hat{\lambda}_1}\E|U-\tilde{U}|^\alpha
+ \frac{1 - \lambda_1^0}{1-\hat\lambda_1}\E|U-\tilde{U}|^\alpha,
\end{equation*}
$$
\frac{\lambda_1^0 - \hat\lambda_1}{1-\hat\lambda_1}-\frac{(\lambda_1^0 -
	\hat\lambda_1)^2}{(1-\hat\lambda_1)^2}- \frac{(1 - \lambda_1^0)(\lambda_1^0
	- \hat\lambda_1)}{(1-\hat\lambda_1)^2} = 0,
$$
and
$$
\frac{1 - \lambda_1^0}{1-\hat\lambda_1}-\frac{(1 - \lambda_1^0)^2}{(1-\hat\lambda_1)^2}-
\frac{(1 - \lambda_1^0)(\lambda_1^0 - \hat\lambda_1)}{(1-\hat\lambda_1)^2}
= 0.
$$
These observations lead to  
\begin{equation*}
\begin{split}
L &= 
\frac{\lambda_1^0 - \hat{\lambda}_1}{1- \hat{\lambda}_1}\mathcal{E}(U,U+X_1'(\beta^c_2-\beta^0_1);\alpha)
\\
&+ 
\frac{1 - \lambda_1^0}{1-\hat\lambda_1}\mathcal{E}(U,U+X_2(\beta^c_2-\beta^0_2);\alpha)\\
&-\frac{(1 - \lambda_1^0)(\lambda_1^0 - \hat\lambda_1)}{(1-\hat\lambda_1)^2}\mathcal{E}(U+X_2'(\beta^c_2-\beta^0_2),U+X_1'(\beta^c_2-\beta^0_1);\alpha).
\end{split}
\end{equation*}
Consequently, it suffices to prove that
\begin{eqnarray}
\label{eq:positive_needed}
\frac{\lambda_1^0 - \hat{\lambda}_1}{1- \hat{\lambda}_1}\mathcal{E}(U,U+X_1'(\beta^c_2-\beta^0_1);\alpha)
+ 
\frac{1 - \lambda_1^0}{1-\hat\lambda_1}\mathcal{E}(U,U+X_2'(\beta^c_2-\beta^0_2);\alpha)&&
\\
-\frac{(1 - \lambda_1^0)(\lambda_1^0 - \hat\lambda_1)}{(1-\hat\lambda_1)^2}\mathcal{E}(U+X_2'(\beta^c_2-\beta^0_2),U+X_1'(\beta^c_2-\beta^0_1);\alpha)
&>& 0.\nonumber
\end{eqnarray}
Since $\mathcal{E}(u,v;\alpha)$ is a metric, triangle inequality implies
that
\begin{equation*}
\begin{split}
\mathcal{E}(U+X_2'(\beta^c_2-\beta^0_2),U+X_1'(\beta^c_2-\beta^0_1);\alpha)
&\leq \mathcal{E}(U,U+X_1'(\beta^c_2-\beta^0_1);\alpha) \\
&+ \mathcal{E}(U+X_2'(\beta^c_2-\beta^0_2),U;\alpha).
\end{split}
\end{equation*}
Furthermore, since $\beta^0_1 \neq \beta^0_2$, we have that at least one
of the terms $\mathcal{E}(U,U+X_1'(\beta^c_2-\beta^0_1);\alpha)$ and $\mathcal{E}(U,U+X_2'(\beta^c_2-\beta^0_2);\alpha)$
is strictly positive.
We also observe that 
$$
\frac{(1 - \lambda_1^0)(\lambda_1^0 - \hat\lambda_1)}{(1-\hat\lambda_1)^2}
\leq \text{min}\left[\frac{\lambda_1^0 - \hat{\lambda}_1}{1- \hat{\lambda}_1},\frac{1
	- \lambda_1^0}{1-\hat\lambda_1}\right],
$$
and the inequality is strict whenever $\lambda_1^0 < 1$ and $\hat{\lambda}_1
< \lambda_1^0$. As \eqref{eq:positive_needed} is trivially valid for $\lambda_1^0
= 1$ or $\hat{\lambda}_1 = \lambda_1^0$, this completes the proof.

\subsection*{A.3. Proof (of Proposition 2)}

The proof is by contradiction. Assume that there exists one or more change
points that are not consistently estimated. In order to prove the statement,
it suffices to find two clusters $U_i$ and $U_j$ such that $d_\alpha(U_j,U_i)
\rightarrow \infty$. Note first that there now exists at least one $j$ 
such that $T^0_{j-1} \leq \hat{T}_{j-1} < \hat{T}_j \leq T^0_j$. Consequently,
there exists at least one cluster $U_j$ such that $\hat\beta_j \rightarrow
\beta^c_j = \beta^0_j$. Similarly, there exists at least one index $i$ such
that the open interval $(\hat{T}_{i-1},\hat{T}_i)$ contains at least one
true change point. Without loss of generality and for notational simplicity,
we assume that $\hat{T}_1 < T^1_0$ and that $(\hat{T}_1, \hat{T}_2)$ contains
$m-1$ true change points $T^0_1,T^0_{2},\ldots,T^0_{m-1}$, where $2\leq m
\leq k+1$. As in the case of a single change point, we obtain a splitting
$I_1 = [1,\hat{T}_1]$, $I_2 = [\hat{T}_1+1,T_1^0]$, $I_3 = [T_1^0+1,T_2^0],\ldots,
I_{m} = [T_{m-2}^0+1,T_{m-1}^0]$, and $I_{m+1} = [T_{m-1}^0+1,\hat{T}_2]$.
Observe that the cluster 
$U_1$ corresponds to the time indexes contained in $I_1$ and that the cluster
$U_2$ corresponds to the time indexes contained in 
$I = \cup_{j=2}^{m+1} I_j$. As in the case of a single change point, we also
observe that $\beta(i) = \beta_1^c = \beta_1^0$ for each $i \in I_1$, and
that $\beta(i) = \beta_2^c$ for each $i \in I$. However, the true values
differ within intervals $I_j$ as, for $i \in I_j$, we have $\beta^0(i) =
\beta_{j-1}^0$. Finally, we denote by $a_j$ the asymptotic proportions giving
the amount
of observations belonging to the intervals $I_j$. That is, we have $|I_j|
\sim a_j T$.

As in the proof of Proposition 1, we set $\hat{u}_i = u_i
- x_i'(\hat{\beta}(i)-\beta^0(i))$, $U_1 = \{\hat{u}_t\}_{t\in I_1}$ and
$U = \{\hat{u}_t\}_{t\in
	I}$. It now suffices to prove that, for some constant $C>0$, we have
\begin{equation}
\label{eq:empirical_limit_multi}
2\hat{\mu}^{\alpha}_{U_1,U}-\hat{\mu}^{\alpha}_{U_1}-\hat{\mu}^{\alpha}_{U}
\rightarrow C,
\end{equation}
where 
$$
\hat{\mu}^{\alpha}_{U_1,U}  =\frac{1}{|I_1||I|}\sum_{j \in I_1, i \in I}
|\hat{u}_i-\hat{u}_j|^{\alpha}, 
$$
$$
\hat{\mu}^{\alpha}_{U_1}  = \frac{1}{|I_1|^2}\sum_{i,j \in I_1}|\hat{u}_i-\hat{u}_j|^{\alpha},
$$
and
$$
\hat{\mu}^{\alpha}_{U}  = \frac{1}{|I|^2}\sum_{i,j\in I}|\hat{u}_i-\hat{u}_j|^{\alpha}.
$$
As the statement given in the step 1 of the proof of Proposition 1
holds for any subsets, we can directly proceed to computing the limits. The
term $\hat{\mu}^{\alpha}_{U_1}$ can be treated as before, and we obtain
$$
\lim_{T\rightarrow\infty} \hat{\mu}^{\alpha}_{U_1} = \E|U-\tilde{U}|^\alpha.
$$
For the term $\hat{\mu}^{\alpha}_{U_1,U}$, we use $I = \cup_{j=2}^{m+1} I_j$.
Now, for each subinterval $I_j$ separately, again by Assumption 5 and Lemma 2, we have that
$$
\lim_{T\rightarrow \infty}\frac{1}{|I_1||I|}\sum_{i\in I_1,j\in I_j}|u_i
- u_j + x_j'(\beta(j)-\beta^0(j))|^\alpha = \frac{a_j}{\sum_{k=2}^{m+1}a_k}\E|U-\tilde{U}+X_{j-1}'(\beta^c_2
- \beta^0_{j-1})|^\alpha.
$$
Thus we obtain
$$
\lim_{T\rightarrow \infty} \hat{\mu}^{\alpha}_{U_1,U} = \sum_{j=2}^{m+1}\frac{a_j}{\sum_{k=2}^{m+1}a_k}\E|U-\tilde{U}+X_{j-1}'(\beta^c_2
- \beta^0_{j-1})|^\alpha.
$$
Similarly, for the last term $\hat{\mu}_{U}^\alpha$, we have
\begin{equation*}
\begin{split}
&\lim_{T\rightarrow \infty}\frac{1}{|I|^2}\sum_{i\in I_{n_1},j\in I_{n_2}}|u_i
- u_j -x_i'(\beta(i)-\beta^0(i)) + x_j'(\beta(j)-\beta^0(j))|^\alpha \\
=& \frac{a_{n_1}a_{n_2}}{\left(\sum_{k=2}^{m+1}a_k\right)^2}
\E|U-\tilde{U} -X_{n_1-1}'(\beta^c_2-\beta^0_{n_1-1}) + \tilde{X}_{n_2-1}'(\beta^c_2-\beta^0_{n_2-1})|^\alpha,
\end{split}
\end{equation*}
leading to
\begin{equation*}
\begin{split}
\lim_{T\rightarrow \infty}\hat{\mu}_{U}^\alpha 
&=\sum_{n_1,n_2=2}^{m+1}\frac{a_{n_1}a_{n_2}}{\left(\sum_{k=2}^{m+1}a_k\right)^2}
\E|U-\tilde{U} -X_{n_1-1}'(\beta^c_2-\beta^0_{n_1-1}) + \tilde{X}_{n_2-1}'(\beta^c_2-\beta^0_{n_2-1})|^\alpha.
\end{split}
\end{equation*}
We proceed as in the step 3 of the proof of Proposition 1.
By using the definition of energy distance, we have that, for any $i$ and
$j$,
\begin{equation*}
\begin{split}
\mathcal{E}(U,U+X_j'(\beta^c_2-\beta^0_j);\alpha) 
&= 2\E|U-\tilde{U}+X_j'(\beta^c_2 - \beta^0_j)|^\alpha - \E|U-\tilde{U}|^\alpha
\\
&- \E|U-\tilde{U} + (\tilde{X}_j'-X_j')(\beta^c_2-\beta^0_j)|^\alpha,
\end{split}
\end{equation*}
and
\begin{equation*}
\begin{split}
\mathcal{E}(U+X_j'(\beta^c_2-\beta^0_j),U+X_i'(\beta^c_2-\beta^0_i);\alpha)&=2\E|U-\tilde{U}+X_i'(\beta^c_2-\beta^0_i)-\tilde{X}_j'(\beta^c_2-\beta^0_j)|^\alpha
\\
&- \E|U-\tilde{U}+(X_i'-\tilde{X}_i')(\beta^c_2-\beta^0_i)|^\alpha\\
& - \E|U-\tilde{U}+(X_j'-\tilde{X}_j')(\beta^c_2-\beta^0_j)|^\alpha.
\end{split}
\end{equation*}
Together with the observation
\begin{equation*}
\E|U-\tilde{U}|^\alpha = \sum_{j=2}^{m+1}\frac{a_j}{\sum_{k=2}^{m+1}a_k}\E|U-\tilde{U}|^\alpha,
\end{equation*}
and
$$
\frac{a_j}{\sum_{k=2}^{m+1}a_k} - \frac{a^2_j}{\left(\sum_{k=2}^{m+1}a_k\right)^2}
- \sum_{l=2,l\neq j}^{m+1}\frac{a_la_j}{\left(\sum_{k=2}^{m+1}a_k\right)^2}
= 0 
$$
this leads to 
\begin{equation*}
\begin{split}
&\lim_{T\rightarrow \infty}\left(\hat{\mu}_{U_1,U}^\alpha - \hat{\mu}_{U_1}^\alpha
- \hat{\mu}_{U}^\alpha\right) \\
&=\sum_{j=2}^{m+1}\frac{a_j}{\sum_{k=2}^{m+1}a_k}\mathcal{E}(U,U+X_{j-1}'(\beta^c_2-\beta^0_{j-1});\alpha)\\
&-\sum_{n_1,n_2=2,n_1\neq n_2}^{m+1}\frac{a_{n_1}a_{n_2}}{2\left(\sum_{k=2}^{m+1}a_k\right)^2}\mathcal{E}(U+X_{n_1-1}'(\beta^c_2-\beta^0_{n_1-1}),U+X_{n_2-1}'(\beta^c_2-\beta^0_{n_2-1});\alpha)
\end{split}
\end{equation*}
which is positive by the arguments given in the step 3 of the proof of Proposition
1.

\section*{Appendix B: Methods used in the simulation study}
In this section, we settings used for the test algorithms and their implementations.
In a \textit{univariate case,} we had 4 benchmarks: NPD, NSA, PSA, BP, ECP. PSA
proposed by~\cite{gorskikh-16}, as well as NPD and NSA were implemented by
us as an R code. BP \citep{bai-perron98,bai-perron03} is available in an
R-package 'strucchange' \citep{Zeileis2001}.
The ideas behind the implementation are described in~\cite{Zeileis2003}.
ECP~\citep{matteson-james14} is implemented in R package 'ecp'.

The settings for these methods were: 
\begin{itemize}
	\item BP: segment length $h$=50
	\item PSA:  $\Theta=5$, $\theta=3$, $\Delta=50$, $\delta=20$ (which
	are correspondingly the number of contributing parameters and the length
	of segments at step 2 and 3)
	\item NSA: $\gamma=0.6$, $l=50$, $s= \tau =50$, $e=T-\tau=550$, $p_0=5\%$
	\item NDP (same as for NSA): $\tau =50$, $p_0=5\%$.
	\item ECP: $sig.lvl=5\%$, $\alpha=1$
\end{itemize}
For all methods, we select the minimum distance between change locations
to be 50, which means that the maximum possible number of breaks detected
will not exceed $T/\tau-1=11$.

In a \textit{multivariate case,} we had 3 competing methods: NPD, NSA and
QP. The first two methods (NPD and NSA) were implemented by us as an R code,
whereas QP \citep{qu-perron-07} is available as a GAUSS code at Pierre Perron's
homepage \url{http://people.bu.edu/perron/}.

The settings for these methods were:
\begin{itemize}
	\item QP: m=11 (number of breaks allowed)
	\item NSA: $\gamma=0.6$, $l=50$, $s= \tau =50$, $e=T-\tau=550$, $p_0=5\%$
	\item NDP (same as for NSA): $\tau =50$, $p_0=5\%$.
\end{itemize}

\newpage

\section*{Appendix C: Application to financial news analytics}

\subsection*{C.1. News-event detection model}

\begin{figure}[!ht]
	\centering
	\includegraphics[width=14cm,height=7cm]{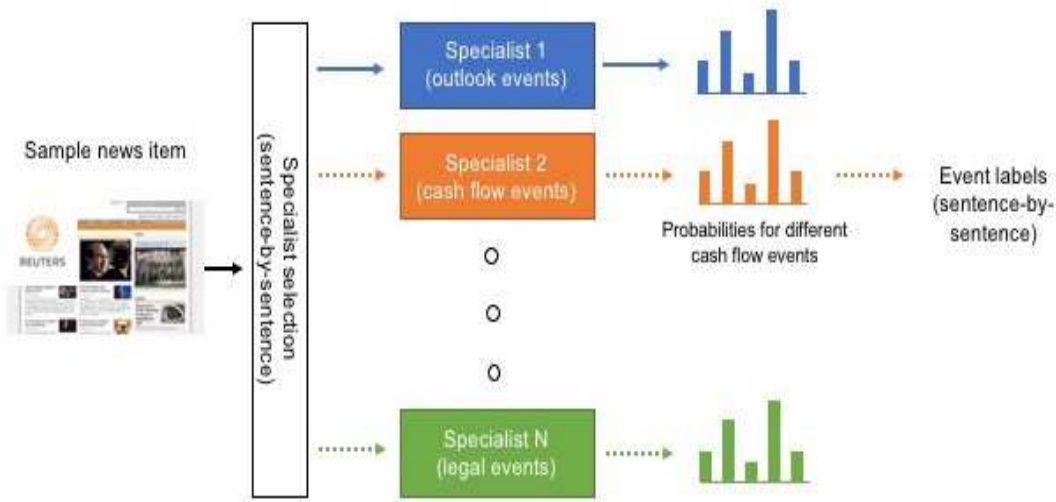}
	\caption{Multi-specialist network for event tagging}\label{fig:multi-specialist-network}
\end{figure}

Fine-grained labeling tasks with thousands of categories are  difficult to
solve using a single classifier due to model capacity constraints and slow
training speed (Ahmed et al. 2016; Gao et al. 2017). A common strategy to
deal with this kind of problem is to divide output tags into semantically
related subgroups (verticals) and train a specialist model per each subgroup
separately. In our financial news analytics case, such strategy is relatively
easy to implement, since there exists a natural taxonomy for organizing 
the events in a semantic hierarchy. For example, all fine-grained {\it legal
	events} can be grouped into one vertical while all {\it outlook events} can
be grouped into another, and so on (see Figure~\ref{fig:multi-specialist-network}).
Each vertical may have a different number of output tags and also different
amounts of training data. The overall model can then be represented as a
tree-structured network with  specialists representing  branches. The choice
of specialist is guided by a selector model (a course category classifier)
that is optimized to discriminate the verticals.  

\begin{figure}[!ht]
	\center
	\includegraphics[width=12cm,height=10cm]{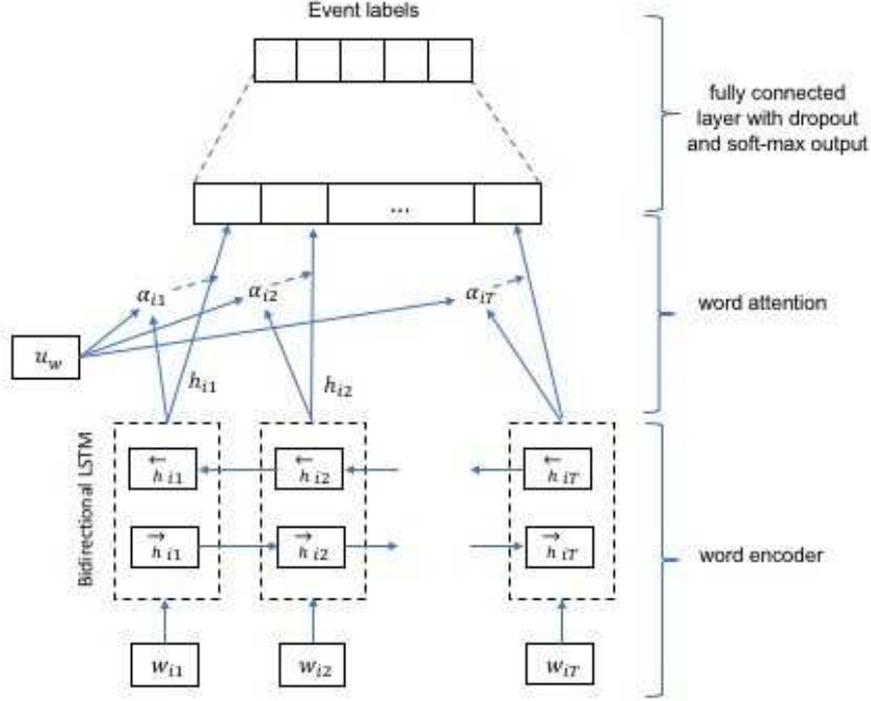}
	\caption{Bidirectional LSTM network with attention mechanism}\label{fig:bi-lstm}
\end{figure}

In our setup, each specialist network as well as the selector are modeled
as bidirectional Long Short-Term Memory (LSTM) networks~\citep{seo-16} with
an attention mechanism (Figure~\ref{fig:bi-lstm}). For simplicity, all event
types considered in this study are assumed to be identifiable from sentence-level
data.
If identification of document-level events is needed, a hierarchical attention\
network can be considered~\citep{yang-16}.   

As described in Figure~\ref{fig:bi-lstm}, given a sentence with words $w_{it}$,
$t\in[0,T]$, we first embed the words to vectors using a pre-trained embedding
matrix $W_e$. The embeddings are then encoded using a standard bidirectional
LSTM layer~\citep{seo-16}:
\begin{eqnarray*}
	x_{it}&=&W_ew_{it}, \ t\in[1,T] \\
	\overrightarrow{h}_{it}&=&\overrightarrow{\text{LSTM}}(x_{it}), \ t\in[1,T]
	\\
	\overleftarrow{h}_{it}&=&\overleftarrow{\text{LSTM}}(x_{it}), \ t\in[T,1]
\end{eqnarray*}
The use of bidirectional LSTM summarizes information from both directions
for words. The contextually enriched word encodings are then obtained by
concatenating the forward and backward hidden states, i.e. $h_{it}=[\overrightarrow{h}_{it},\overleftarrow{h}_{it}]$.
To extract words that are most relevant for the identifying the events in
the sentence, this is followed by simple word attention mechanism~\citep{yang-16}
to compute importance weighted encodings. The normalized importance weights
$\alpha_{it}$ are given by
\begin{eqnarray*}
	u_{it}&=&\text{tanh}(W_e h_{it} + b_w) \\
	\alpha_{it}&=&\frac{\exp(u^T_{it}u_w)}{\sum_t \exp(u^T_{it}u_w)},
\end{eqnarray*}
where $u_{it}$ is a hidden representation of $h_{it}$. As a final stage,
the importance weighted word encodings are then passed to a fully connected
layer with dropout and soft-max activation, which will then compute the probabilities
for different event labels.

\subsection*{C.2. Bank-specific analysis of structural change points}

\begin{figure}[!htp]
	\centering

	\includegraphics[width=14cm,height=13.9cm]{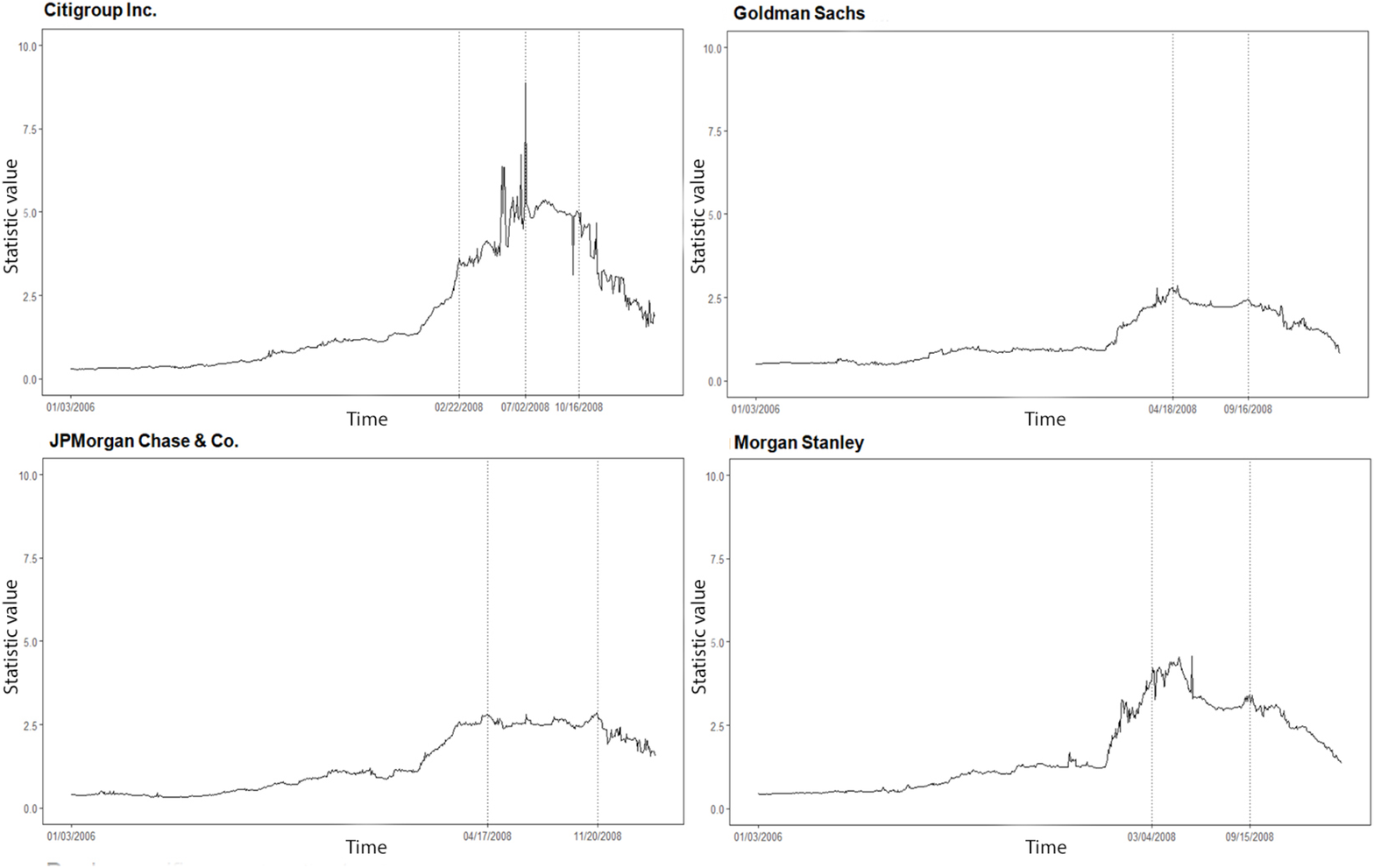}
	
	\caption{\small Bank-specific energy-distance statistics and structural
		change points. The dashed vertical lines indicate the locations of structural
		change points detected using bootstrap test statistics.}\label{fig:bank-specific}
\end{figure}

\begin{figure}[!htp]
	\centering
	
	\includegraphics[width=16cm,height=10cm]{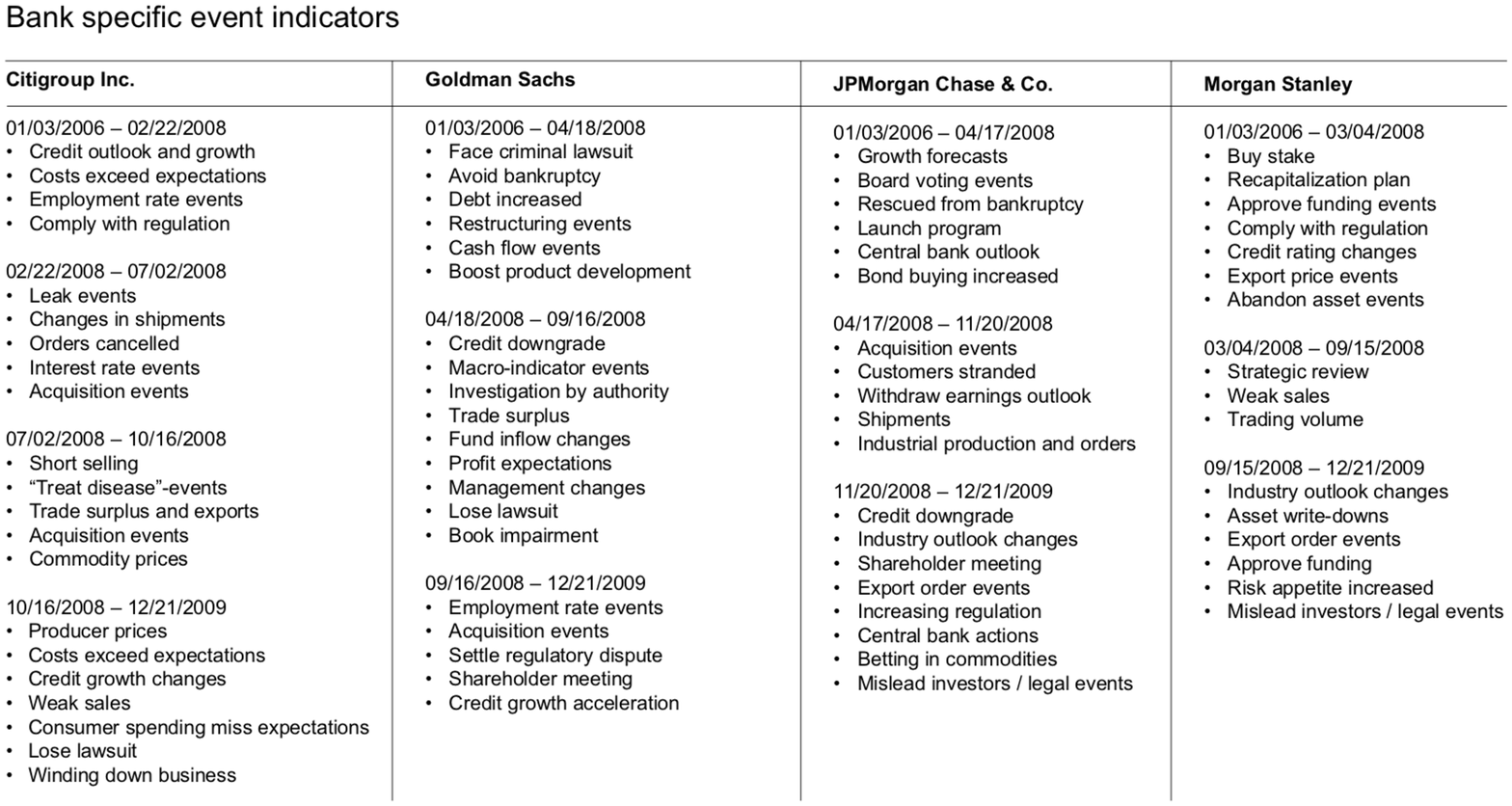}
	\caption{\small Collection of bank-specific event-type variables
		by regime.  }\label{fig:banks-events}
\end{figure} 

Figures~\ref{fig:bank-specific} and~\ref{fig:banks-events} provide more details
on the regimes from the perspective of the individual banks. Notably, the
general shape of the energy distance graphs in Figure~\ref{fig:bank-specific}
is relatively similar, and the variation in the number and length of regimes
looks modest. For convenience, we show the energy-statistics only four banks, since the graphs of the remaining banks are very similar. In general, it looks like 2 or 3 structural change points are found. The multivariate statistic suggests 3 change points, which are located in the middle of May-2007, May-2008 and August-2008. When considering the statistics for the individual banks, we see a bit more variation in number and location of changes, but they are,
nevertheless, quite close to the ones detected by the multivariate statistic. However, reflecting the unique state of each bank and the underlying
dynamics of the economy, the subset of contributing event-indicators varies
considerably across regimes as well as banks. Although, the overall number
of possible event types was over 500 in the news wire dataset, the use of
Lasso-regularization lead to rather sparse models with only 5-10 variables
in each; see Figure~\ref{fig:banks-events}. When considering the event types
by regime, it appears that they agree quite well with the ones found by the
multivariate statistic. However, in addition to macroeconomic events there
are quite a lot of company specific legal issues, regulatory disputes, and
news dealing with their restructuring and recapitalization plans.

\end{document}